\documentclass{pasj01}

\begin{document}
\SetRunningHead{Makishima et al.}{Precession in 4U 0142+61}

\Received{2018/7/28}
\Accepted{2018/10/22}

\title{
A NuSTAR study of the 55 ks hard X-ray pulse-phase modulation
in the magnetar 4U~0142+61
}

\author{
Kazuo  \textsc{Makishima},\altaffilmark{1,2,3}
Hiroaki \textsc{Murakami},\altaffilmark{2}
Teruaki \textsc{Enoto},\altaffilmark{4}\\
and
Kazuhiro \textsc{Nakazawa}\altaffilmark{2,5}
}

\email{maxima@phys.s.u-tokyo.ac.jp}

\altaffiltext{1}{
High Energy Astrophysics Laboratory, and MAXI Team, 
The Institute of Physics and Chemical Research (RIKEN),
2-1 Hirosawa, Wako, Saitama 351-0198
}
\altaffiltext{2}{
Department of Physics, The University of Tokyo,
7-3-1 Hongo, Bunkyo-ku, Tokyo 113-0033
}
\altaffiltext{3}{
Kavli Institute for the Physics and Mathematics of the Universe,
The University of Tokyo,
5-1-5 Kashiwa-no-ha, Kashiwa, Chiba, 
277-8683
}
\altaffiltext{4}{
The Hakubi Center for Advanced Research, Kyoto University, Kyoto 606-8302,
and 
Department of Astronomy, Kyoto University, 
Kitashirakawa-Oiwake-cho, 
Sakyo-ku, Kyoto 606-8502
}
\altaffiltext{5}{
Division of Particle and Astrophysical Science, and
Kobayashi-Masukawa Institute for the Origin of Particles and the Universe, Nagoya University,
Furo-cho Chikusa-ku, Nagoya 464-8602\\
}

%
\maketitle

\begin{abstract}
Archival NuSTAR data of the magnetar 4U 0142+61,
acquired in 2014 March for a total time span of 258 ks,  were analyzed.
This is to reconfirm the 55 ks  modulation in the hard X-ray  pulse phases  of this source,
found with a Suzaku observation in 2009 (Makishima et al. 2014).
Indeed, the 10--70 keV X-ray pulsation,
detected with NuSTAR at  $8.68917$ s,
was  found to be also phase-modulated  (at $> 98\%$ confidence)
at  the same $\sim 55$ ks period, or half that value.
Furthermore, a brief analysis of another Suzaku data set of 4U 0142+61,
acquired in 2013,  reconfirmed the same 55 ks
phase modulation in the 15--40 keV pulses.
Thus, the hard X-ray  pulse-phase modulation  was detected
with Suzaku (in 2009 and 2013) and NuSTAR (in 2014) at a consistent period.
However, the modulation amplitude varied significantly;
$A \sim 0.7 $ s with Suzaku (2009), $A \sim 1.2 $ s with Suzaku (2013), 
and $A \sim 0.17 $ s with NuSTAR.
In addition, the phase modulation properties detected with NuSTAR
differed considerably between the  first 1/3 and the latter 2/3 of the observation.
In  energies below 10 keV,
the pulse-phase modulation was not detected
with either Suzaku or NuSTAR.
These results reinforce the view of Makishima et al. (2014);
the neutron star in  4U 0142+61 keeps free precession,
under a slight axial deformation  due probably to ultra-high 
toroidal magnetic fields of $\sim 10^{16}$ G.
The wobbling angle  of precession should remain constant,
but the pulse-phase modulation amplitude varies on time scales of months to years,
presumably as asymmetry of the hard X-ray emission pattern 
around the star's axis changes.

\end{abstract}


\section{Introduction}
\label{sec:intro}
Magnetars  (e.g., \cite{Magnetar,Mereghetti08}), a particular subclass of neutron stars (NSs),
are considered to harbor ultra-strong magnetic fields (MFs) 
and to emit X-rays by consuming their huge magnetic energies.
While conducting  magnetar studies  \citep{Enoto10,Enoto17}
with the Suzaku X-ray observatory,
we found an intriguing effect in the persistently bright magnetar 4U 0142+61
(\cite{Makishima14}, hereafter Paper I):
in its observation made in 2009 August for 
a gross exposure of 186 ks (net 102 ks),
the source's pulsation detected with the Hard X-ray Detector 
(HXD; \cite{HXD1,HXD2}) at a period of 8.69 s was
found to suffer periodic slow phase modulation, with a period of 
\begin{equation}
T=55 \pm 4~{\rm ks}
\label{eq:55ks}
\end{equation}
and a modulation amplitude of $A \sim 0.7$ s,
in the 15--40 keV energy band
where the hard spectral component (hereafter HXC;  \cite{Enoto10,Enoto11}) dominates.
In fact, the HXC pulsation became detectable 
only after applying a {\it demodulation} correction to the data.
A follow-up Suzaku observation made in 2013
successfully reconfirmed the hard X-ray phase modulation,
with a possibly larger amplitude of $A\sim 1.2$ (see section \ref{subsec:Suzaku0913}).
The effect was however apparently absent (Paper I)
in an earlier Suzaku observation made in 2007 August,
under nearly the same exposure and same source intensity 
as in 2009 and 2013.
Furthermore, in energies below $\sim 10$ keV
where the soft spectral component (SXC) dominates,
the 55 ks modulation was  absent  on all these  occasions,
with a typical upper limit of $A<0.3$ s.

In Paper I (see figure 3b therein) and \citet{Max16},
we interpreted the above Suzaku results  in the  following way.
\begin{enumerate}
\item 
The NS in this system is axially symmetric;
its moment of inertia $I_3$ around the symmetry axis $\hat{x}_3$
(to be identified with the magnetic dipole axis)
differs slightly from that  in orthogonal directions, $I_1$.
The deformation is specified by  {\it asphericity}, defined as
 \begin{equation}
  \epsilon \equiv (I_1 - I_3)/I_3~.
  \label{eq:epsilon}
  \end{equation}
\item
The NS, being axisymmetric,  undergoes free precession.
The $\hat{x}_3$ axis rotates around  the constant angular momentum $\vec{L}$
with the precession period $P_{\rm pr} \equiv 2\pi I_1/L$,
keeping a constant  {\it wobbling} angle $\alpha$  to $\vec{L}$.
At the same time, the NS rotates around $\hat{x}_3$
with the rotation period $P_{\rm rot} \equiv 2\pi I_3/L$.
The observed pulse period should be identified with $P_{\rm pr}$ rather than $P_{\rm rot}$.
 \item
Beat between $P_{\rm pr}$ and $P_{\rm rot}$
defines a long period at
\begin{equation}
T \equiv  \left[(1/P_{\rm rot} - 1/P_{\rm pr}) \cos \alpha \right]^{-1} 
 = P_{\rm pr}/( \epsilon \cos \alpha) ~,
\label{eq:slip}
\end{equation}
which is called {\it slip period}.
Substituting  equation (\ref{eq:55ks}) for this $T$,
and assuming $\cos \alpha \approx 1$,
we obtain 
 \begin{equation}
  \epsilon =P_{\rm pr}/(T \cos \alpha) \approx P_{\rm pr}/T =1.6 \times 10^{-4}~.
  \label{eq:epsilon_value}
  \end{equation}
\item 
If the emission pattern is axially symmetric around $\hat{x}_3$,
we would detect periodicity at
neither $P_{\rm rot}$  nor $T$.
This condition is thought to apply to the  SXC pulsation,
of which the pulse phase is not modulated. 
\item
If the X-ray emission pattern, in contrast,  breaks symmetry around $\hat{x}_3$,
the  pulse arrival times (and hence the pulse phase) become modulated at $T$.
This is thought to explain the behavior of the HXC pulses.
If the degree of this asymmetry varies,
the modulation amplitude $A$ will differ among observations,
even though $\alpha$ should remain constant.

\item 
We regard the deformation as  prolate ($\epsilon >0$),
in which case the free precession should develop spontaneously.
Such a prolate deformation as equation (\ref{eq:epsilon_value}) can 
in turn be caused by strong toroidal MF  $B_{\rm t}$ hidden inside the NS, as
 \begin{equation}
  \epsilon  \sim 1\times 10^{-4} (B_{\rm t}/10^{16}{\rm G})^2~
  \label{eq:deformation}
  \end{equation}
\citep{Cutler02,Ioka+Sasaki04}.
Hence the NS in 4U 0142+61 is suggested to harbor $B_{\rm t} \sim 10^{16}$ G,
which exceeds by two orders of magnitude its nominal dipole MF of
$B_{\rm d}= 1.3\times 10^{14}$ G \citep{4U0142_Bd}.
%
%
\end{enumerate}

The Suzaku discovery deserves further studies,
because the effect potentially provides vital  information
on the  internal MF of magnetars,
which is otherwise difficult to measure.
Actually, we have discovered a second example,
the highly variable and fastetst-rotating magnetar 1E~1547.0$-$5408.
In this case, the Suzaku data obtained in its 2009 January outburst revealed
that the hard X-ray pulses with  $P_{\rm pr} =2.0721$ s
are phase-modulated at a period of $T=36$ ks  (\cite{Makishima16}, hereafter Paper II).
Again assuming $\cos \alpha \sim 1$,
equation (\ref{eq:slip}) yields $\epsilon= 0.6 \times 10^{-4}$,
which is of the same order as equation (\ref{eq:epsilon_value})  for 4U~0142+61.

Since the above  results have all been obtained with Suzaku,
we evidently need to study this intriguing phenomenon 
using other X-ray observatories.
Actually, \citet{TEA15}, hereafter TEA15,  
searched the NuSTAR data of 4U 0142+61 acquired in 2014 March
for the same phase-modulation effects,
and found no evidence over a range of $T=45-65$ ks.
However,  as already suggested by the multiple Suzaku data (above item 5),
$A$ would vary  as the anisotropy of the HXC emission pattern changes
(with $\alpha$ kept constant).
It is hence important to re-examine the NuSTAR data
for evidence of pulse-phase modulation,
at a period consistent with equation~(\ref{eq:55ks}),
but allowing for different values of $A$.
With these in mind, 
we re-analyze the same NuSTAR data as TEA15 used.

\section{Observation }

As described by TEA15,
4U 0142+61 was observed with NuSTAR in 2014 March in two successive pointings.
The first shorter one (ObsID 30001023002) was
from 2014 March 27 UT 13:35 to March 28 UT 00:45,
for a gross exposure of 41 ks and a net exposure of 24  ks.
The 2nd and longer one (ObsID 30001023003) 
immediately followed the 1st one, 
and lasted from  March 28 UT 00:45 to March 30 UT 13:01.
The achieved gross and net exposures are 217 and 144 ks, respectively.

In the present work,
the two data sets are co-added  for our analysis,
to cover an overall time span of 258 ks (a net exposure of 168 ks).
The events taken with the two focal-plan modules, 
FPMA and FPMB, are also co-added together.
We accumulated on-source events within a circle of radius $60''$
around the image centroid,
because the signal-to-noise ratio of the source
was maximized at a radius of $60''$ to $70''$,
in the 20--70 keV range which plays an important role in our study.
Then, subtracting the background, 
and applying dead-time and vignetting corrections,
the object was detected at a 4--70 keV count rate of 
$1.009\pm 0.002$ c s$^{-1}$ for FPMA+FPMB.
Finally, like in TEA15, the event arrival times were photon-by-photon 
converted to those to be measured at the solar-system barycenter;
this utilized  the source position of $(\alpha^{2000}, \delta^{2000})
= (01^{\rm h}46^{\rm m}22^{\rm s}.41,  61^\circ45'03.''2)$,
as well as  the spacecraft orbital information.

\section{Data Analysis and Results}
%
  
\subsection{Basic analyses}
\label{subsec:basic_ana}

\subsubsection{Light curves}
\label{subsubsec:light_curves}

\begin{figure}[tbh]  
  \begin{center}
   \FigureFile(80mm,){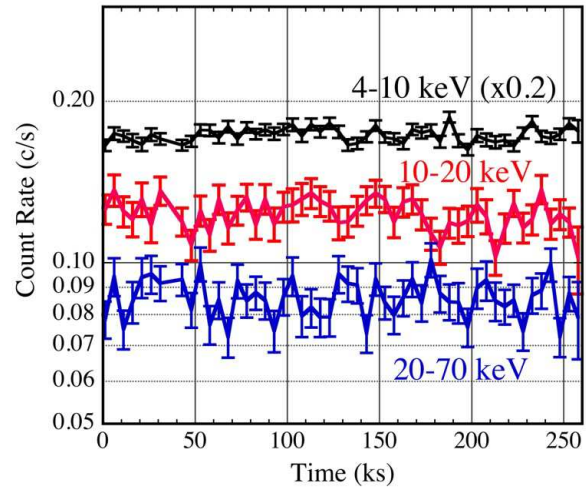}
\end{center}
     \caption{Background-inclusive light curves in the 4--10 keV
     (L; black), 10--20 keV (M; red), and 20--70 keV (H; blue) bands,
      all binned into 5 ks. 
  Error bars are $\pm 1 \sigma$.
The L-band data are multiplied by a factor of 0.2 for presentation.
}
\label{fig:LC_5ks}
\end{figure}

Figure~\ref{fig:LC_5ks} shows 
dead-time-corrected but background-inclusive 
light curves of 4U 0142+61 with  5 ks binning.
Here and hereafter, 
the overall energy range is subdivided into three;
4--10 keV, 10--20 keV, and 20--70 keV,
to be called L, M, and H bands, respectively.
The energy range below 4 keV (down to the nominal lower bound of 3 keV)
was discarded to avoid any possible drifts
in the detector's lower threshold.
Similarly, we discard the highest energy interval of 70--78 keV,
because of the reduced effective area therein.
The L band covers the SXC,
whereas M and H mainly represent the HXC.
 Using the $60''$ event accumulation radius,
the background  amounts to approximately
1.2\%, 5.2\%, and 32\% of the total counts
in  L, M, and H, respectively.

In figure ~\ref{fig:LC_5ks},
the assumption of a constant count rate gives
 $\chi^2/\nu=1.52$ , 1.21, and 1.12,
 in the L, M, and H bands, respectively, all for $\nu=50$.
Therefore, some intrinsic intensity variations are suggested
in the L band (with a null hypothesis probability of 1\%),
and possibly in the M band (7.5\%) as well,
but  the H band variations are dominated by Poisson fluctuations.
 
\subsubsection{Hardness ratios}
\label{subsubsec:hardness_ratios}

\begin{figure}[tbh]  
  \begin{center}
     \FigureFile(83mm,){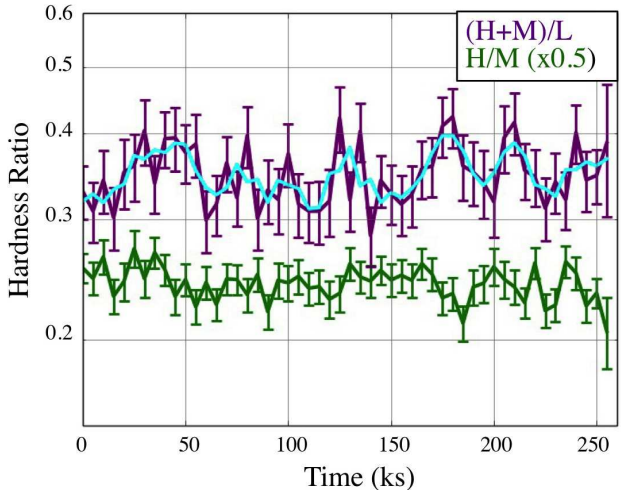}
     \end{center}
 \medskip
     \caption{Time histories of the two hardness ratios, 
     (M+H)/L (green) and M/H (purple), with $\pm 1 \sigma$ error bars.
  The former is halved for presentation.
 The cyan curve superposed on the latter shows
 the result of 3-bin running average.
    }
\label{fig:HR_5ks}
\end{figure}

In figure~\ref{fig:LC_5ks},
the M and H count rates show a hint of  anti-correlation,
even though their intrinsic variations are 
not so significant (section \ref{subsubsec:light_curves}).
This is suggestive of changes in the HXC slope.
We hence produced in figure~\ref{fig:HR_5ks}
time histories of two hardness ratios, (M+H)/L and H/M.
The former, i.e., the HXC vs. SXC intensity ratio shown in green,
is found to be consistent with being constant within errors.
The latter, representing the HXC slope,
in contrast exhibits more noticeable variations.
Actually, after applying the 3-bin running average
(25\%, 50\%, and 25\% weights for consecutive 3 data points),
it yields $\chi^2/\nu \sim 2.0$ ($\nu \sim 25$) 
against the constant-value hypothesis,
with an implied null-hypothesis probability of $\sim 0.2\%$.
Here, we considered that the employed running average
approximately halves both the total noise power
and the effective degrees of freedom.
Thus, the HXC slope exhibits statistically significant variation,
in agreement with the   M vs. L anti-correlation 
suggested  by figure~\ref{fig:LC_5ks}.

\begin{figure}[htb]  
  \begin{center}
  \FigureFile(82mm,){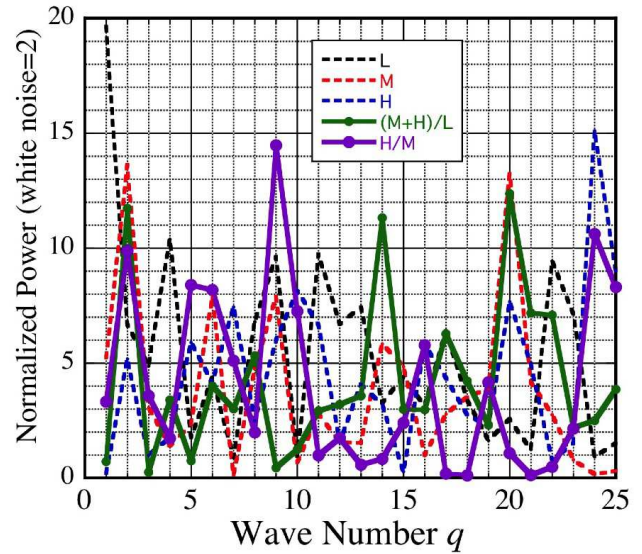}
   \end{center}
  \caption{
Fourier power spectra, calculated from the three light curves in 
 figure~\ref{fig:LC_5ks} and the two hardness ratios in figure~\ref{fig:HR_5ks}.
 They are  all normalized so that the power becomes 2.0
 when the statistical white noise dominates.
 Their colors are the same as in figure~\ref{fig:LC_5ks}  and figure~\ref{fig:HR_5ks}.
 The wavenumber $q$ has a period of $258/q$ ks.
   }
 \label{fig:HR_periodicity}
  \end{figure}

The  variations in the H/M ratio seen in 
figure~\ref{fig:HR_5ks}  are not only significant, 
but also appear rather periodic, with a period of 5--6 bins (25--30 ks).
We hence  Fourier-analyzed the two hardness ratios in figure~\ref{fig:HR_5ks},
as well as the three-band light curves in figure~\ref{fig:LC_5ks}.
The obtained five power spectra
are presented in  figure~\ref{fig:HR_periodicity}.
Thus, in the H/M power spectrum (purple),
the power reaches $\sim 14.48$ 
at a wavenumber of $q=9$, or a period of $\sim 258/q  \sim 28.7$ ks
which  is  consistent wit just half the period of equation (\ref{eq:55ks}).
In addition, some power enhancements are seen at $q \sim 5$
(period of $\sim 52$ ks) which could be a subharmonic.
To evaluate the  significance of the $q=9$  peak, we must consider the fact 
that the average of the H/M power spectrum (excluding $k=0$) reaches 4.14,
which is 2.07 times higher than the value of 2.0 
expected for a purely Poisson-dominated case.
Therefore, the source is varying in the H/M ratio.
Regarding the value of 4.14
as a quadratic sum of the Poissonian fluctuations and this intrinsic source variation,
and assuming the latter to have a white spectrum,
we  renormalized the power at $q=9$ to $14.48/2.07=7.00$,
to obtain a chance  probability of 1.5\% for this peak.
The probability becomes rather high, $\sim 38\%$,
when  multiplied by the overall wave numbers of 25
to take into account so-called ``look-else where effects".
However, if we limit our search to the 1st two (fundamental and the 2nd) harmonics
and the 1/2 sub harmonic of equation (\ref{eq:55ks}),
the probability still remains $\sim 4.5\%$,
and hence the peak is significant  at a confidence  of $\sim 90\%$ or higher.
A related  discussion on this point is further given at section~\ref{subsec:significance}.

We hence conclude 
that the HXC of 4U 0142+61 observed with NuSTAR exhibits
some evidence of spectral-slope variations at a period
that  is consistent with half the period of equation (\ref{eq:55ks}).

\subsubsection{Pulsations}
\label{subsubsec:pulsations}
In the present study, 
we search the data for periodicities using the  $Z_N^2$ analysis 
(\cite{Zn2_83,Zn2_94,Enoto11}; Paper I; Paper II;
see also section~\ref{subsubsec:demod_proc}).
By applying this method 
to the background-inclusive 10--70 keV  events,
with the harmonic number set at $N=3$ (see  below),
the 8.69 sec source pulsation was detected throughout the observation.
Since the local pulse period obtained 
from the first 6 ks (gross) of data 
could be inconsistent (shorter) with those from the other intervals,
we exclude, unless otherwise stated,
this first 6 ks from the subsequent data analysis throughout the present study,
although the origin of this effect is unclear.
Our major results remain unchanged
even if this time range is retained.

\begin{figure}[htb]  
  \begin{center}
  \FigureFile(80mm,){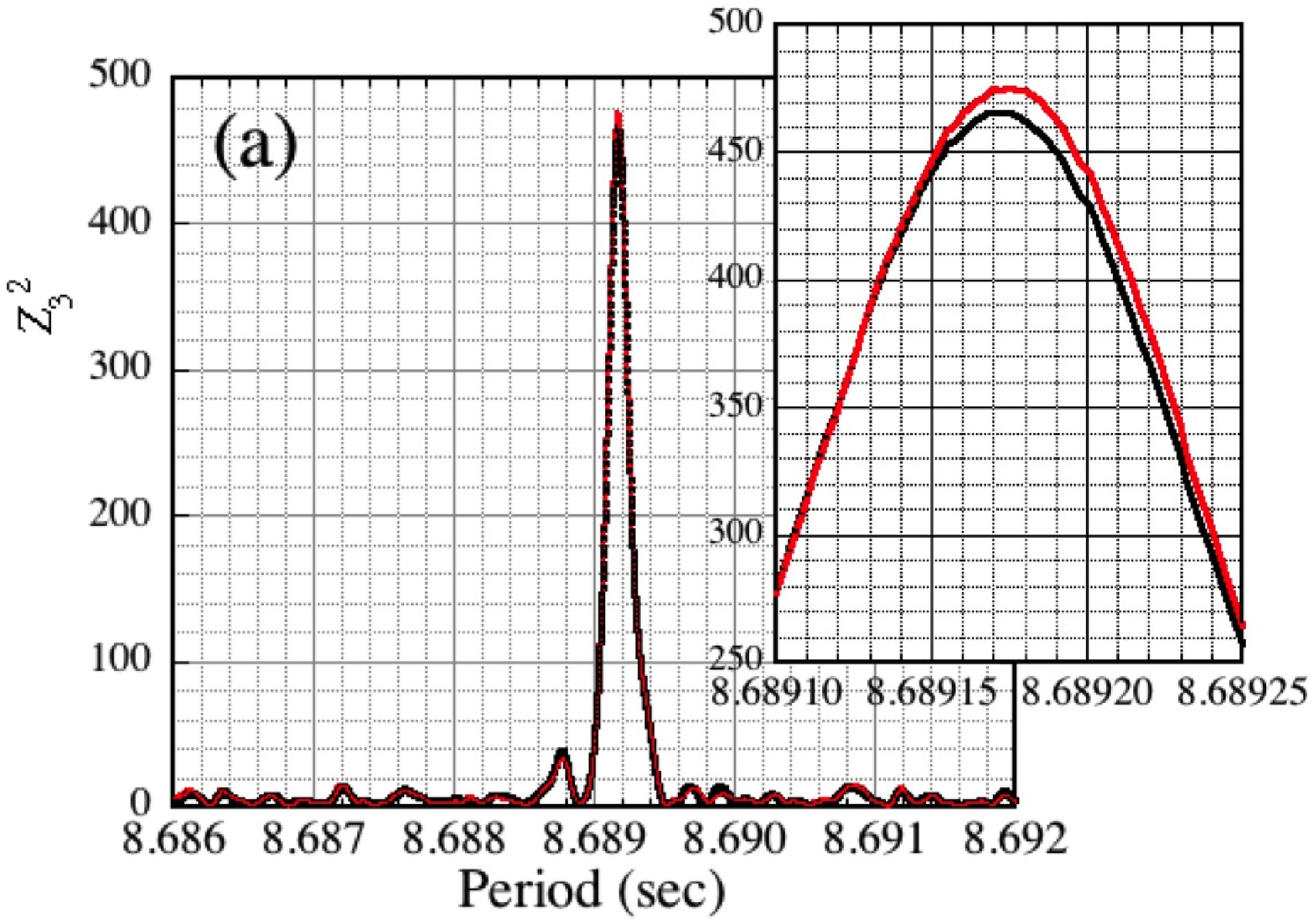}
  \FigureFile(70mm,) {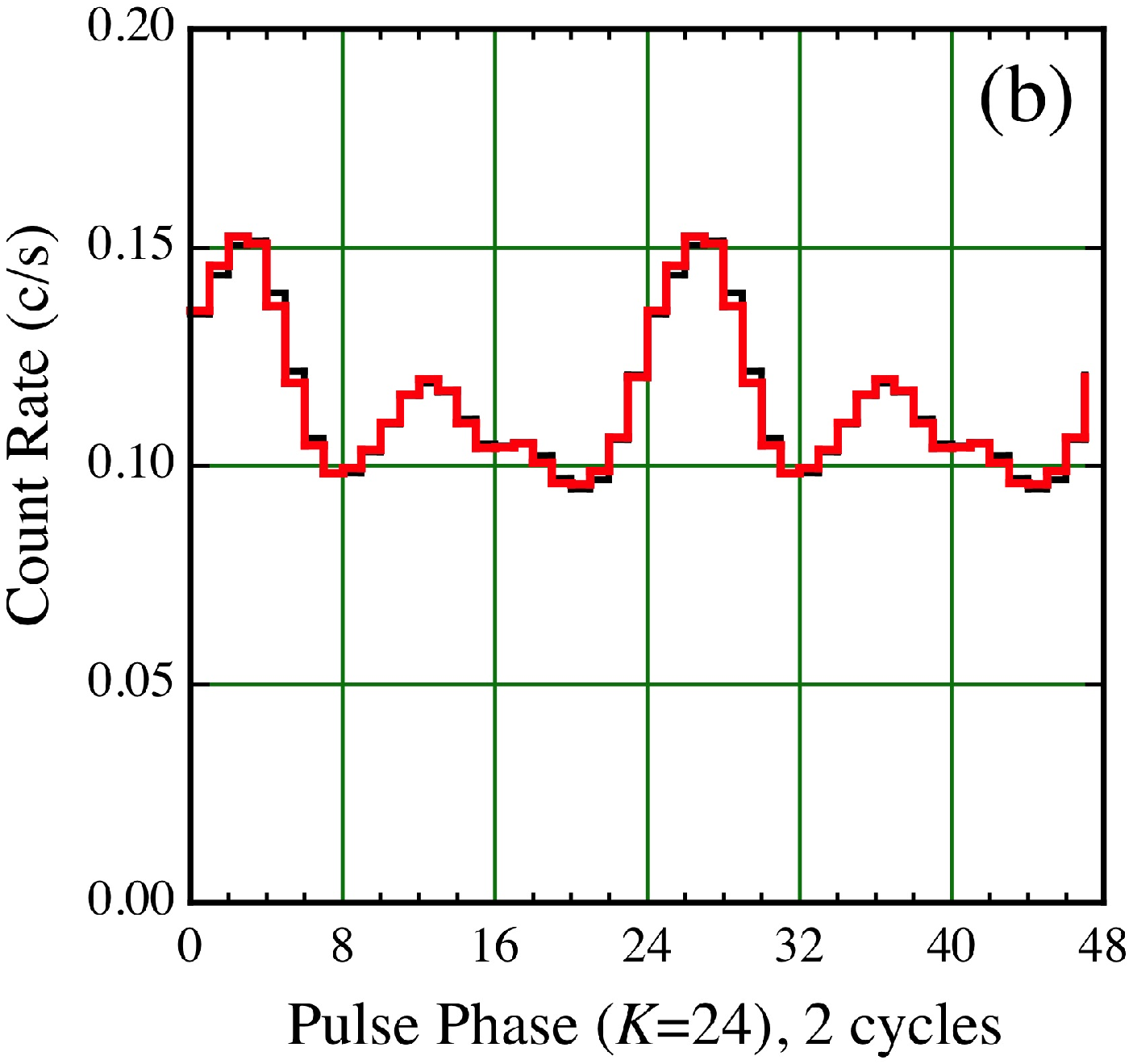}
 \end{center}
  \caption{(a) A $Z_3^2$ pulse periodogram of 4U 0142+61,
  calculated using the background-inclusive 10--70 keV NuSTAR data.
  Black and red are  before and after the demodulation analysis,
 and the inset gives a blow-up around the peak.
  (b) The folded 10--70 keV pulse profile,
  before (black) and after (red) the demodulation,
  smoothed with 3-bin running average.
  The background is inclusive,
  and the count rate is  corrected for neither deadtime nor vignetting.
  The statistical 1-$\sigma$ error is about  $\pm 0.003$ c s$^{-1}$.
   }
 \label{fig:PGPr_DemNodem}
  \end{figure}

An (M+H)-band  pulse periodogram around the pulse period,
obtained from the remaining $252~(=258 - 6)$ ks of gross exposure,
is presented in figure~\ref{fig:PGPr_DemNodem}a.
(Inclusion of the L band gives consistent results.)
The pulsation has thus been detected with a high significance,
at a barycentric period of
\begin{equation}
P_0 = 8.68917 \pm 0.00002~{\rm s}
\label{eq:P0}
\end{equation}
which is consistent with that reported in TEA15.

The black trace in figure~\ref{fig:PGPr_DemNodem}b
shows the background-inclusive 10--70 keV pulse profile, 
folded at equation (\ref{eq:P0}).
Here and hereafter, the present work employs the time origin
as 2014 March 27,  UT 13:36:18.551,
before the barycentric correction is applied.
The profile is double-peaked, with the primary and secondary peaks,
in agreement with TEA15.
Since the two peaks are separated in phase by about 0.37 cycle,  rather than 0.5,
the profile must include harmonics higher than $N=2$.
In fact,  by Fourier transforming the profile,
we found that the power in the  fundamental ($N=1$),
the 2nd ($N=2$), the 3rd, and the 4th harmonics
carry about 32, 53, 9, and 2 percent of the total power 
(including the Poissonian noise), respectively.
Therefore, below we basically adopt $N=3$, 
even though $N=4$ was used in Paper I.

\subsection{Demodulation analysis}
\label{subsec:demod}

The study  has now come to the stage of  examining
whether the periodicity of equation (\ref{eq:55ks}),
found with Suzaku, is also present in the NuSTAR data.
For this purpose,  we apply
the  {\it demodulation analysis}, developed in Paper I and Paper II,
to the NuSTAR data.

\subsubsection{Procedure}
\label{subsubsec:demod_proc}

The demodulation analysis assumes
that  the  arrival time  $t$ of each pulse peak is modulated sinusoidally
for some unspecified causes by
\begin{equation}
\delta t = A \sin (2\pi t/T -\psi_0)~,
\label{eq:modulation}
\end{equation}
with respect to the case of an exactly constant pulse intervals.
Here $T$, $A$,  and $\psi_0$ are the period, amplitude, 
and the initial phase of the assumed modulation, respectively.
Then, by shifting the arrival times of individual photons  back by $-\delta t$,
we re-calculate $Z_N^2$ for various values of $T$,
to see whether  the pulse significance increases.
More specifically,
this  analysis is carried out in the following way (see Paper II).

First, we select a trial value of $T$, 
and sort the 10--70 keV events (including background)
into a two-dimensional array $C(j, k)$;
here, $j=1, 2, .., J-1 $ specifies the phase in $T$,
$J$ is the total phase number in $T$,
$k=1, 2, .., K-1$ is the pulse phase variable (i.e., the time modulo $P$),
and $K$ is  the overall pulse-phase bin number.
After an appropriate exposure correction, 
$\{C(j, k); k=0, 1, .., K-1\}$  gives the
folded pulse profile at the $j$-th phase of the period $T$.
This procedure may be  called {\it double folding}.

Next, for a pair of parameters ($A, \psi_0$), 
a {\it demodulated} pulse profile is produced as 
\begin{equation}
 D(k) = \sum_{j=0}^J  C\left(j,k'(j) \right)
  \label{eq:Z2_demod}
\end{equation}
where the pulse phase is systematically modulated as
\begin{equation}
 k'(j) = k + K  (A/P)  \cdot \sin( 2\pi j/J - \psi_0 )~.
\label{eq:Z2_wiggle}
\end{equation}
Equation (\ref{eq:Z2_demod}) reduces to the ordinary folding procedure,
if putting $A=0$, or equivalently,  $k'=k$.

The profile $D(k)$ is then Fourier transformed,
and the obtained power is summed up to the specified harmonic number $N$.
The $Z^2$  statistic is calculated by normalizing the summed power in a certain way,
 to reflect signal-to-noise ratios of the data \citep{Zn2_83,Zn2_94}.
Finally, we scan over $T$, $A$, and $\psi_0$,
searching for the triplet $(T, A, \psi_0)$
that maximizes $Z^2$ through equation (\ref{eq:Z2_demod}) and equation (\ref{eq:Z2_wiggle}).

\begin{figure*}[htb]
  \begin{center}
  \FigureFile(90mm,){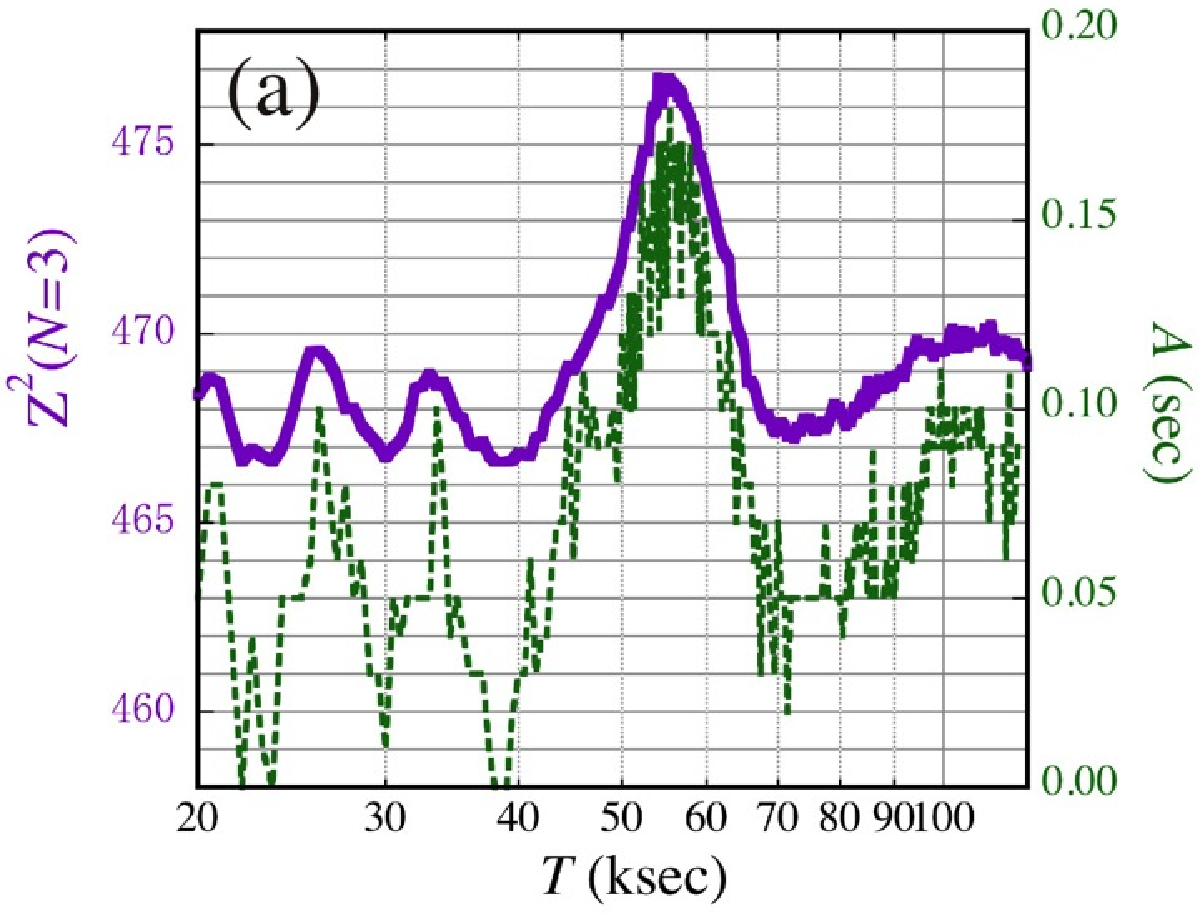}
 \hspace{2mm}
  \FigureFile(60mm,){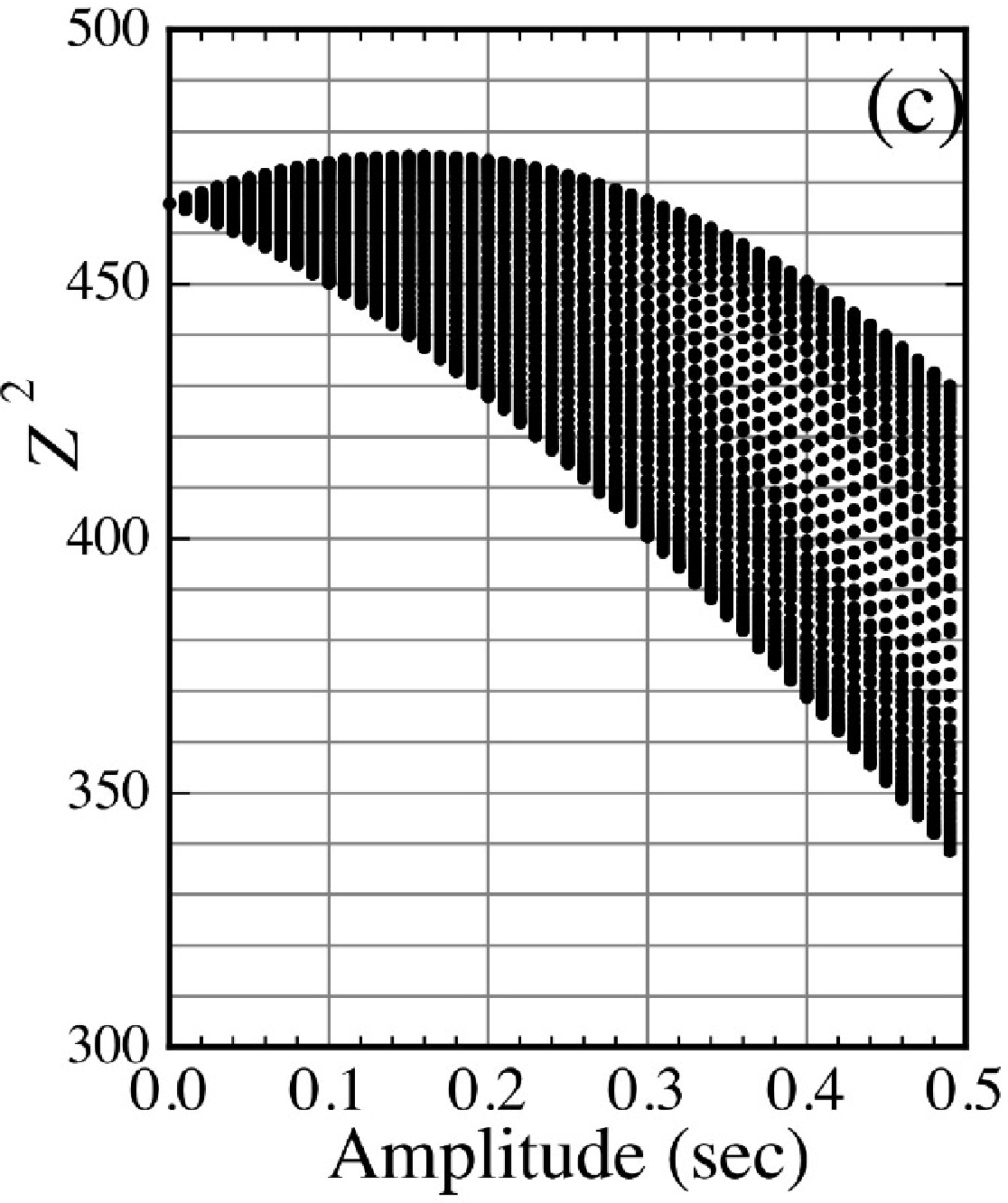}
 \hspace*{5mm}
  \FigureFile(70mm,){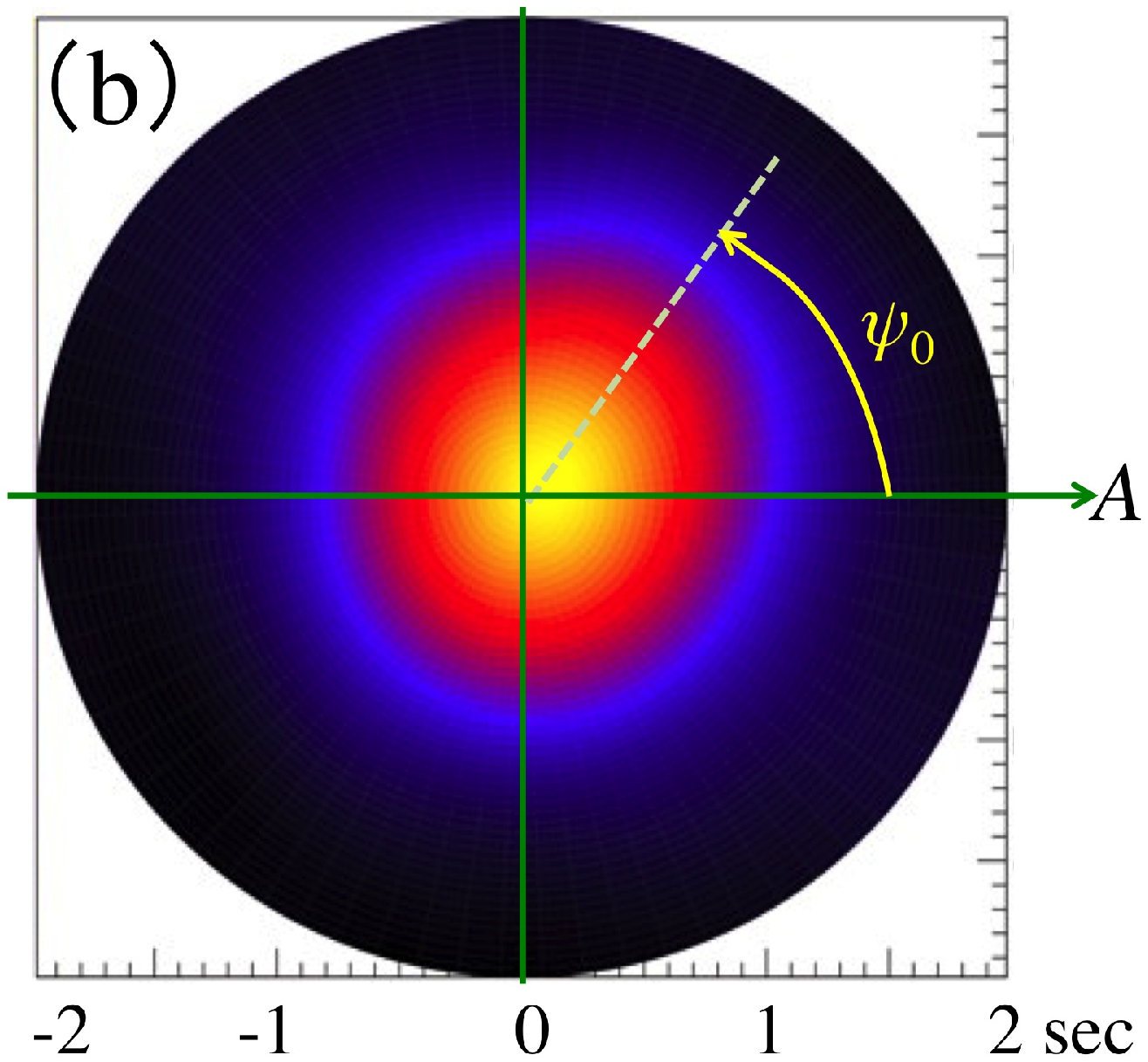}
 \hspace{5mm}
  \FigureFile(70mm,){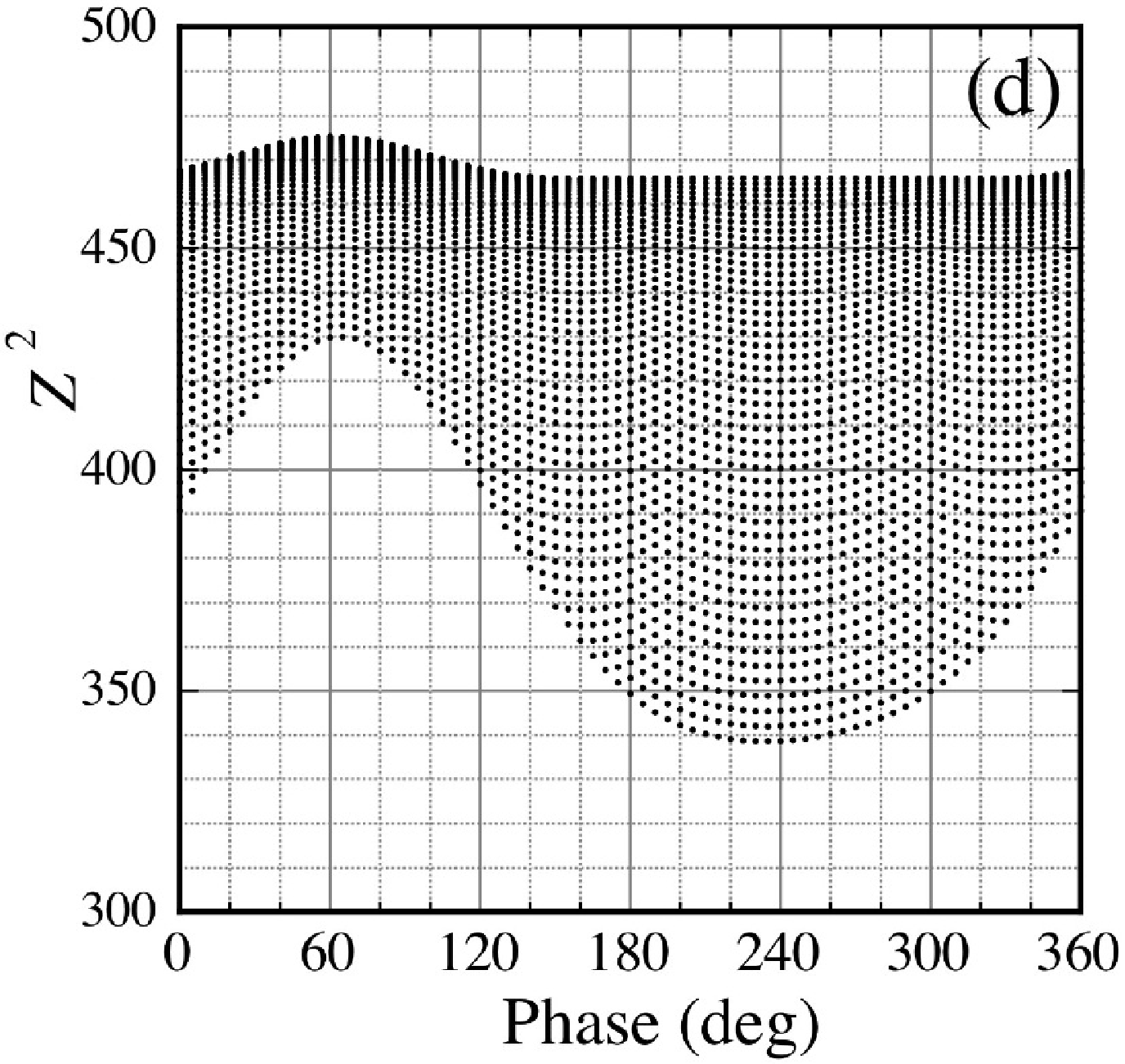}
  \end{center}
  \caption{Results of the demodulation analysis 
  on the 10--70 eV (M+H) NuSTAR data using $N=3$,
  where $Z_3^2$ is abbreviated as $Z^2$.
  (a) The maximum value of $Z^2$ (purple, ordinate to the left)  found at each $T$.
  The dashed green trace (ordinate to the right) shows the values of $A$
  that maximized $Z^2$.
 (b) A color-scale presentation of $Z^2$ on the $(A, \psi_0)$ polar coordinates,
 for a particular period of $T=55$ ks.
 The phase $\psi_0$ is measured anti-clockwise from due right.
 (c) The projection of panel (b) onto the $A$ axis,
 which is shown by a green arrow in panel (b).
 (d) That on the phase ($\psi_0$) axis,
 which is indicated in panel (b) in yellow.
   }
 \label{fig:demodulation}
  \end{figure*}

\subsubsection{Results in 10--70 keV}
\label{subsubsec:demod_results}

The analysis  described above  
has been applied to the 10--70 keV NuSTAR data,
where the search scans were performed
over $T=20-120$ ks with a step of 0.2--2 ks,
$A=0-0.25$ s with a $0.01$ s step, 
and  $\psi_0=0-360^\circ$ with a $5^\circ-10^\circ$ step;
$P$ was scanned over the error range of equation (\ref{eq:P0}) with a  $5~\mu$s step.
We employed  the $Z^2$ harmonic number of  $N=3$ 
(section~\ref{subsubsec:pulsations}),
and large values as  $J=90$ and $K=360$
to make the results in equation (\ref{eq:Z2_demod}) insensitive to $J$ and $K$.
Unless ambiguous, $Z_N^2$ is hereafter abbreviated as $Z^2$.

The derived results are summarized in figure~\ref{fig:demodulation},
which is the same plot as figure 2 of Paper I and figure 4 of Paper II
except some stylistic changes.
The solid purple curve in panel (a),
to be called a {\it demodulation  periodogram}
(to be distinguish from pulse periodograms
such as figure~\ref{fig:PGPr_DemNodem}a)
displays  the maximum  value of $Z^2$, obtained at each $T$
while scanning in $P$, $A$, and $\psi_0$ over the specified ranges.
At $T \sim 55$ ks which is of our prime interest,
we indeed observe a clear peak.
As indicated by dashed green trace in the same panel (a),
the modulation amplitude that maximizes $Z^2$
also increases to $A\sim 0.17$ s around this modulation period.
These results suggest 
that the pulse-phase modulation effects detected 
with Suzaku are also present in the NuSTAR data,
although the amplitude is much reduced.

The $Z^2$ peak in figure~\ref{fig:demodulation}a
has a full width at half maximum of about $\sim 13$ ks,
which is reasonable as a coherent periodicity 
lasting throughout the observation.
This is because the adjacent independent (Fourier orthogonal)
periods should in general be mutually separated  by
\begin{equation}
\Delta T \sim  T^2/T_{\rm tot}~,
\label{eq:DeltaT}
\end{equation}
where $T_{\rm tot}$ is the total data span.
For $T_{\rm tot}=252$  ks and $T=55$ ks,
this yields $\Delta T =12$ ks.

In figure~\ref{fig:demodulation}a,
we also observe a few smaller $Z^2$ enhancements,
including those at $T\sim 110$ ks and $T\sim 27$ ks,
which are just twice and half the  55 ks period, respectively.
They are also accompanied by some increases in $A$.
Hereafter, the 55 ks and 27 ks humps seen in these 
demodulation periodograms are called Zp55 and  Zp27, respectively.

The significance of  Zp55 and  Zp27 may be represented 
by the relative $Z^2$ increase, defined as 
\begin{equation}
\Delta Z^2 \equiv Z^2 - (Z^2)_0~,
\label{eq:deltaZ2}
\end{equation}
where  $(Z^2)_0$ is the value obtained without demodulation.
Table~\ref{tbl:Zp55Zp27} summarizes these quantities for Zp55 and Zp27.
The statistical significances of these enhancements are evaluated later.

\begin{figure*}[]
  \begin{center}
  \vspace*{-3mm}
\includegraphics[height=5.0cm]{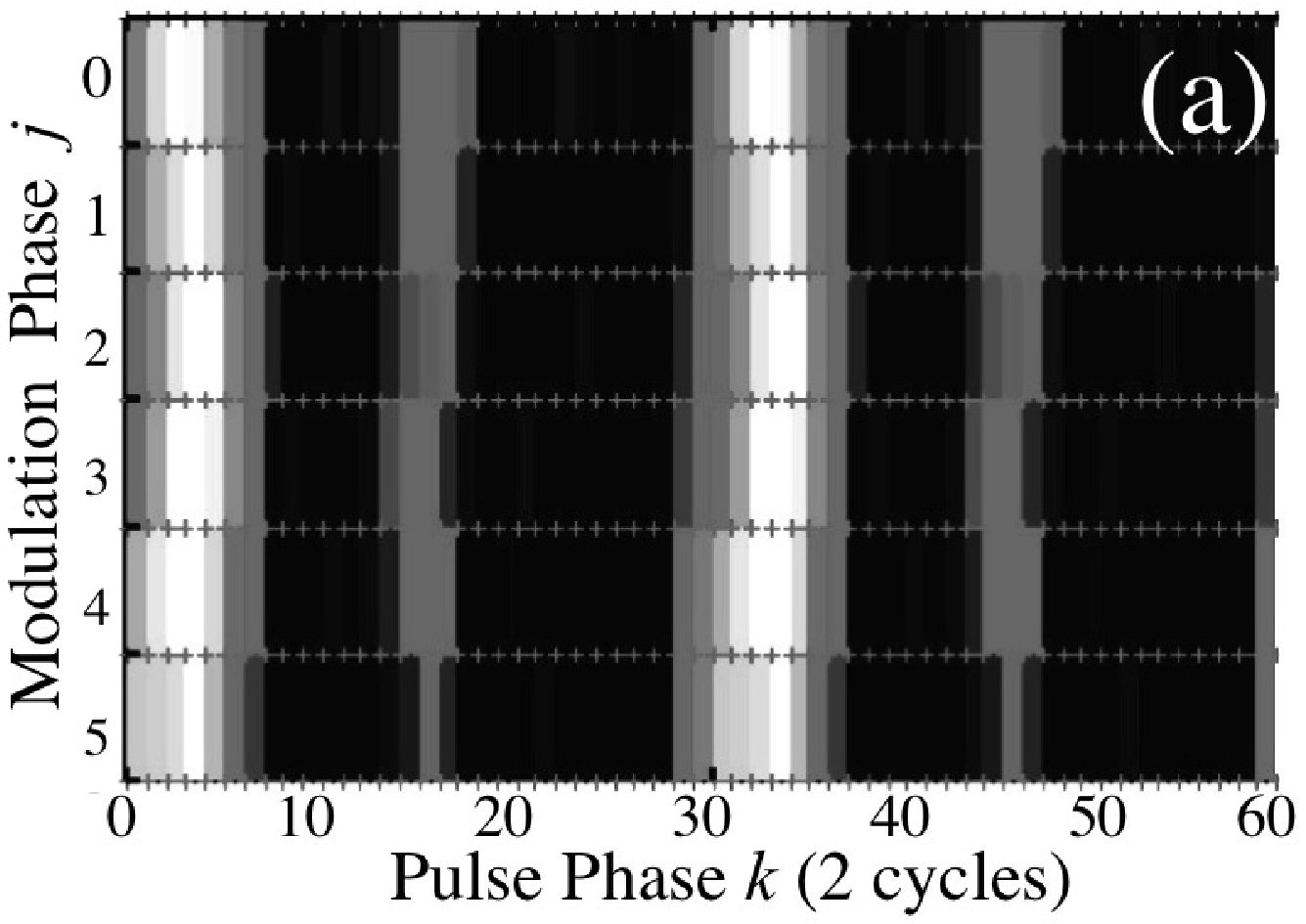}
\hspace{40mm}
\includegraphics[height=6.2cm,]{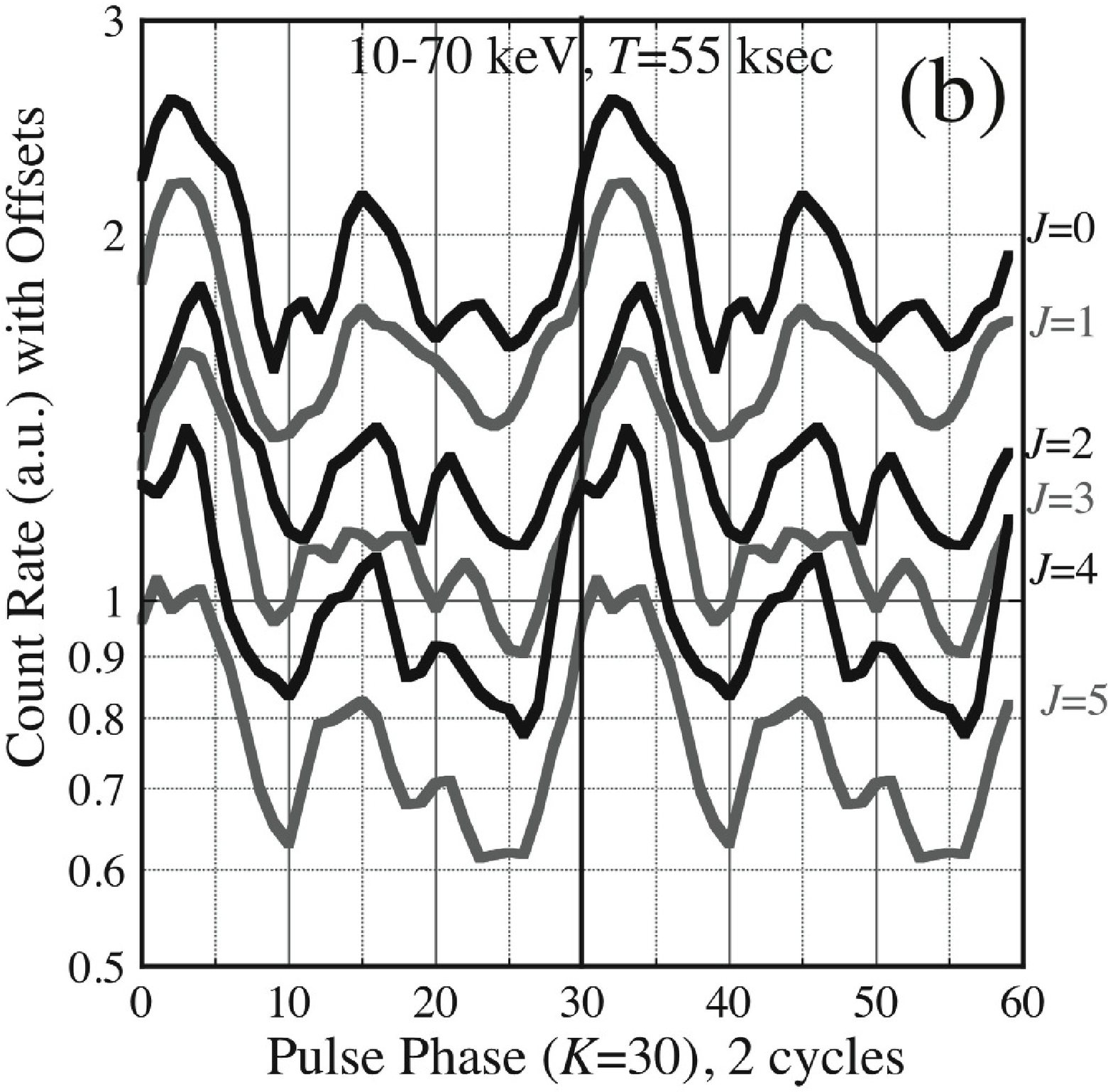}
\includegraphics[height=6.2cm,]{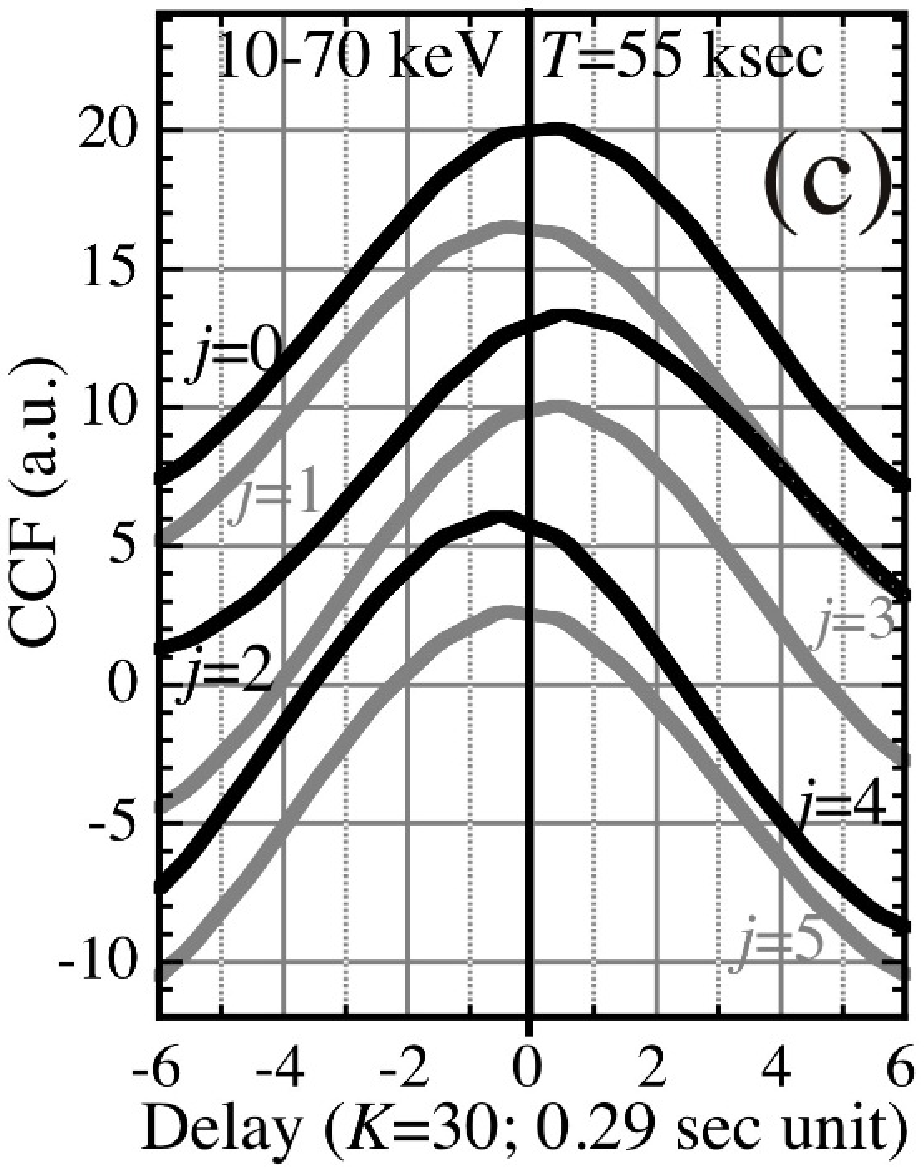}
 \end{center}
  \caption{
  Visualization of the 10--70 keV pulse-phase modulation.
  (a)  A gray-scale count-rate map  $C(j,k)$ obtained by 
  double-folding the 10--70 keV data using $T=55$ ks, 
  $P_0$ of equation (\ref{eq:55ks}), $J=6$, and $K=30$.
  Abscissa shows the pulse phase $k$ for 2 cycles, 
  and ordinate represents the 55 ks modulation phase $j$,
  from top to bottom,  for one cycle.
 (b) The 10--70 keV folded pulse profiles for different $j$,
 shown in semi-logarithmic scale with vertical offsets.
 (c) CCFs of the pulse profiles for $j=0, 1, ,,.  5$, from top to bottom.
 They were calculated as described in text,
  and are shown in arbitrary units with vertical offsets.
  Abscissa is the time lag in units of $P_0/30=0.29$ s,
  where positive values mean delays behind the template.
   }
 \label{fig:visual}
  \end{figure*}

Figure \ref{fig:demodulation}b gives, in a color scale, 
the 10--70 keV $Z^2$ value (the maximum as $P$ is varied)
at each point on the $(A, \psi_0)$ plane,
with $T= 55$ ks  fixed.
We chose a polar-coordinate presentation, instead of the Cartesian form 
which was used in Paper I and Paper II.
There, the pulse-phase modulation effect at $T=55$ ks is visible
as a small offset of the peak position from the coordinate origin ($A=0$).
Projections of the map onto the $A$ and $\psi_0$ axes
are given in panel (c) and panel (d), respectively.
Just for presentation, the range of $A$ in panels (b) and (c) 
is made wider than was used in the actual search ($A<0.25$ s).
Thus,  the parameters characterizing Zp55 are estimated as 
\begin{equation}
T=54.8\pm 5.3~{\rm ks}, ~~ A=0.17\pm 0.08 ~{\rm s}, ~~\psi_0=60^\circ\pm 30^\circ.
\label{eq:demodpara}
\end{equation}
Here,  the errors in $T$ and $\psi_0$
were determined as the standard deviation of a Gaussian
that approximates the peak
when it is added on top of a uniform background (Paper I, Paper II).
Because the $A$ vs $Z^2$ curve (figure~\ref{fig:demodulation}c) 
cannot be described by a constant plus a Gaussian,
we define its error range as the points 
where $\Delta Z^2$  falls $0.607 = {\rm e}^{-0.5}$ times the peak value,
as would be the case with a Gaussian at $\pm 1$-sigma off the centroid.
These parameters are reproduced in the top row of table~\ref{tbl:Zp55Zp27}.

When the demodulation is carried out,
the 10--70 keV pulse  periodogram and the folded pulse profile 
changed as shown in red in figure~\ref{fig:PGPr_DemNodem}a 
and figure~\ref{fig:PGPr_DemNodem}b, respectively.
The difference between ``before" and ``after"  is however very small,
reflecting the rather small increment $\Delta Z^2$.

The demodulation analysis was repeated by chaining $N$ from 3 upwards.
Generally, the $\Delta Z_N^2$ value for Zp55 did not change very much.
However, that for Zp27  increased up to $\Delta Z_{10}^2 \sim 8$ at $N \sim 10$,
beyond which the Poisson noise started to affect the  demodulation  periodogram.
This intriguing effect is further studied in section~\ref{subsec:timediv}.
As a representative cases of higher $N$, 
the parameters derived with $N=7$ are given 
in the second row of table~\ref{tbl:Zp55Zp27}.

\subsubsection{Visualization of the pulse-phase modulation}
\label{subsubsec:demod_visual}
Even though Zp55 is observed rather clearly in figure~\ref{fig:demodulation}a,
the rather  small increment in $Z^2$, namely, $\Delta Z^2 =10.89 \ll (Z^2)_0$,
might cast some doubts on its reality.
Thus, to visualize the suggested effect,
figure \ref{fig:visual}a shows the 10--70 keV double-folding
map $C(j,k)$ (section~\ref{subsubsec:demod_proc}) in a gray scale.
It was produced using  $T_{\rm tot}= 252$ ks,
$T=55$ ks,  and $P_0$ of equation (\ref{eq:P0}),
and is presented with $J=6$ and $K=30$
after applying the three-bin running average 
to the $k$ (abscissa) dimension. 
We have chosen these particular values of $K$ and $J$  
as a compromise between sufficient statistics 
and adequate time resolution.
There, the double-peaked pulse profile (figure~\ref{fig:PGPr_DemNodem}b)
appears as two ridges per cycle, running almost vertically.
In addition, the phase modulation is observed 
as a slight wiggling of the primary (brighter) ridge by $\pm  1$ bin,
corresponding to $A\sim \pm P_0/30 =\pm 0.29$ s.
This map is to be compared with the soft X-ray result
in figure~\ref{fig:demodSXC} to be presented later,
wherein no pulse-phase modulation is observed at $T \sim 55$ ks.

Figure~\ref{fig:visual}b shows the pulse profiles
$\{C(j,k): k=0, 1,2,.., \}$ at individual modulation phases $j$;
it is just another way of presenting panel (a).
To make the primary and secondary peaks both visible,
ordinate utilizes a logarithmic scale.
We reconfirm that the primary pulse peak is modulated,
as a function of $j$,
by approximately the same amount as in panel (a).

At each $j$, we cross-correlated the profile in panel (b),
against an average pulse profile (summed over $j$)
which is shown in red in figure~\ref{fig:PGPr_DemNodem}b.
The same $J=6$ and $K=30$ as panels (a) and (b) were used.
The obtained cross-correlation functions (CCFs) are presented in figure~\ref{fig:visual}c,
where the smoothing method described in Paper I was incorporated.
Again, the correlation peak moves back and forth,
by $\sim \pm 0.5$ unit in this presentation, or $A \sim \pm 0.15$ s,
which  is consistent with equation (\ref{eq:demodpara}).
However, it is smaller than the amplitude,  $A \sim 0.3$ s, 
visible in figure~\ref{fig:visual}a and figure~\ref{fig:visual}b.
This issue is further considered in section \ref{subsec:TwoPeaks}.


\begin{figure}[htb]
  \begin{center}
\includegraphics[height=6.3cm,]{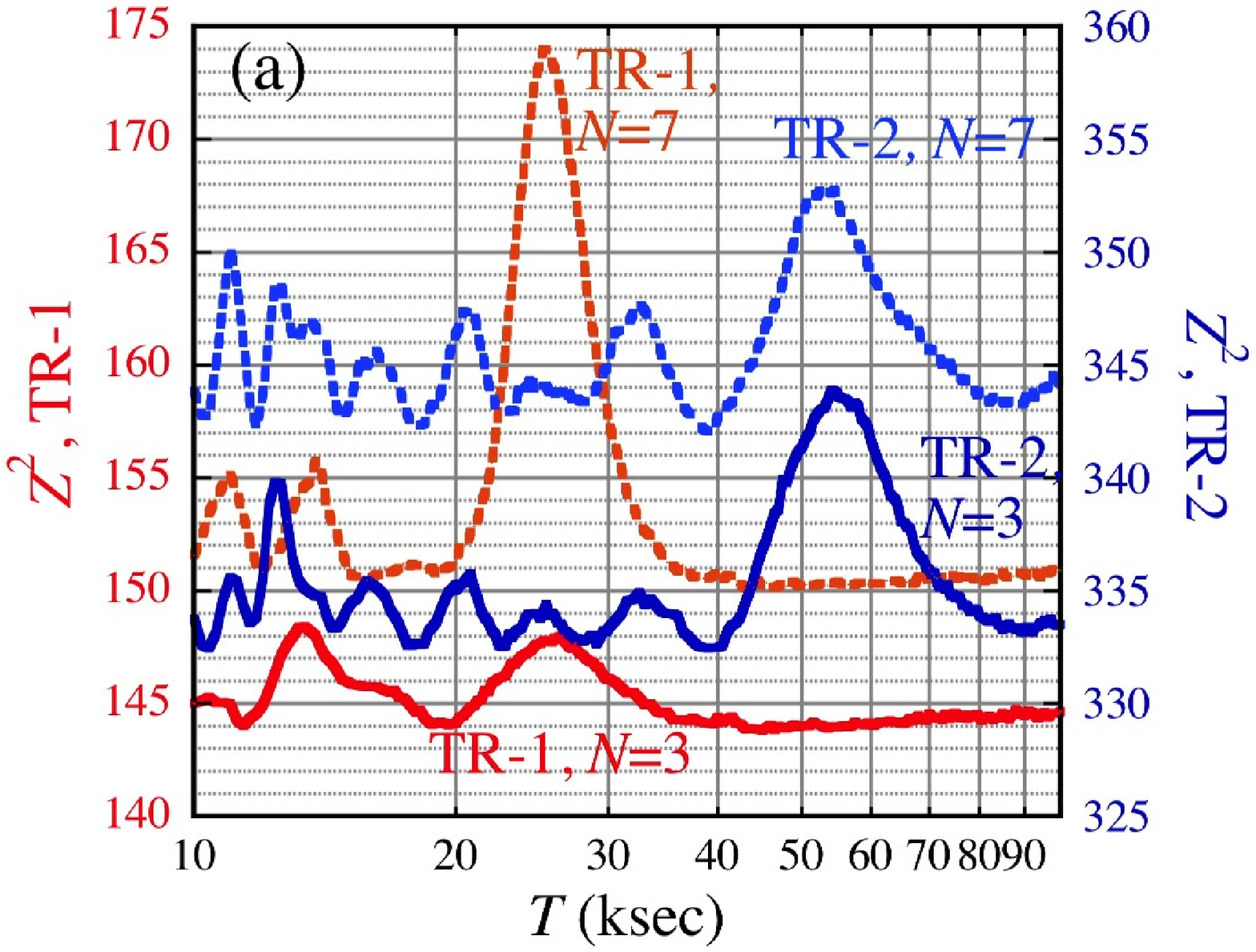}
\hspace{3mm}
\includegraphics[height=5.5cm,]{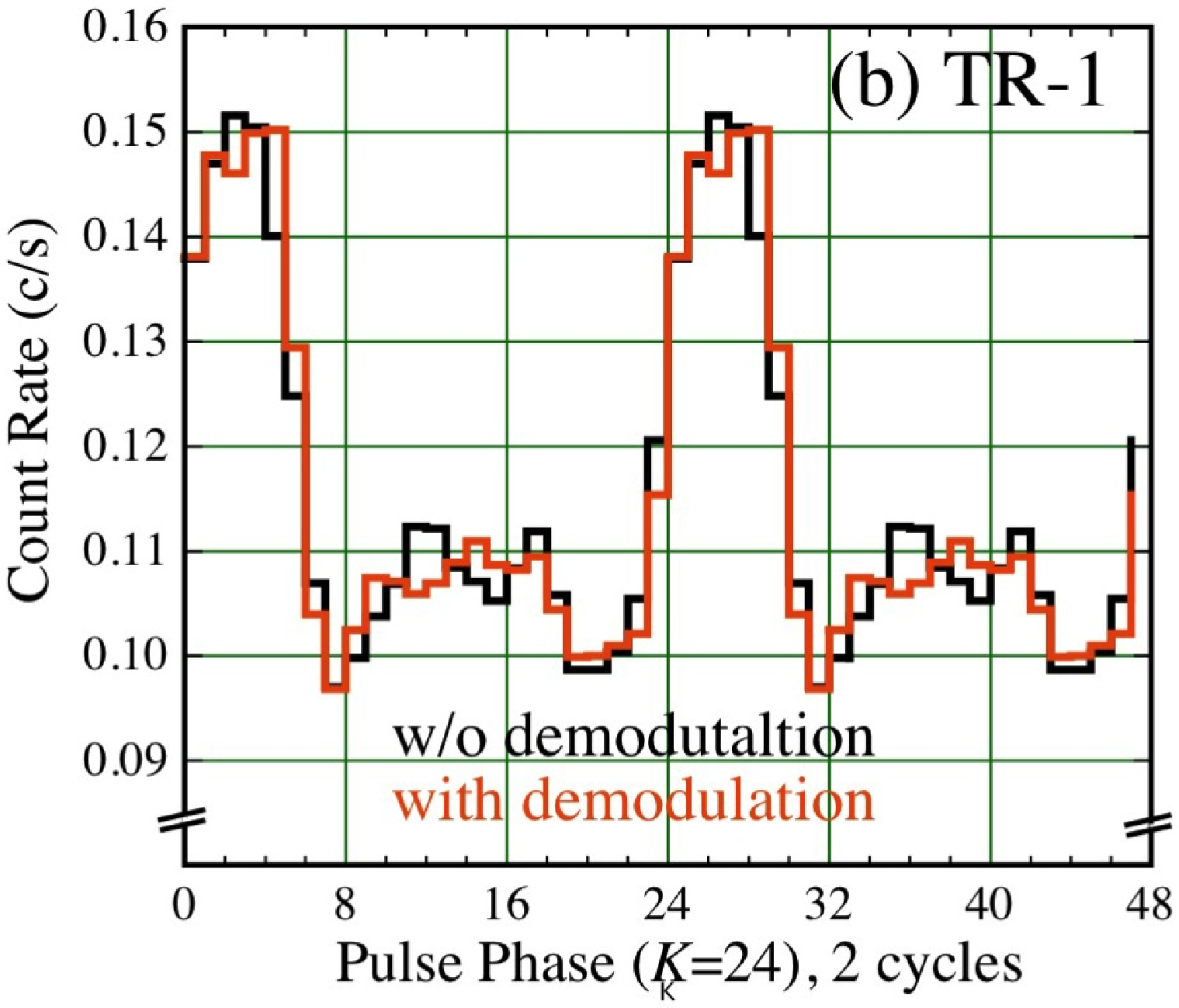}
\includegraphics[height=5.5cm,]{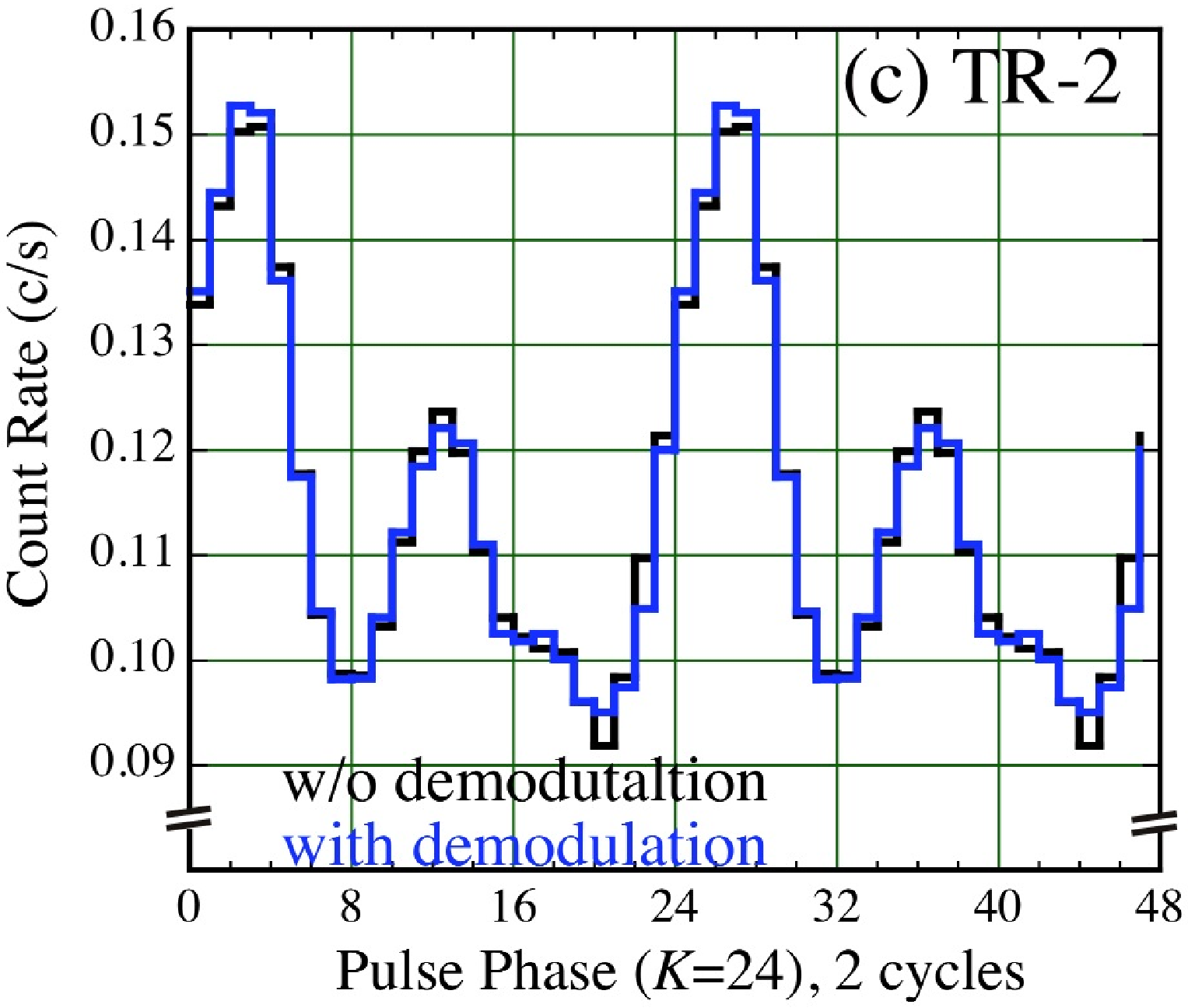}
 \end{center}
  \caption{
  Demodulation analyses applied to the two time ranges,  TR-1 and TR-2.
 (a) The 10--70 keV demodulation periodograms from TR-1,
 with $N=3$ (solid red) and $N=7$ (dashed orange),
 and those from TR-2, 
 with $N=3$ (solid dark blue) and $N=7$ (dashed cyan).
 (b) The folded 10--70 keV pulse profiles from TR-1.
The black one is before demodulation,
whereas the red one is after demodulation 
using the corresponding parameters given in table~\ref{tbl:Zp55Zp27}
with $T=55$ ks.
Unlike figure~\ref{fig:PGPr_DemNodem},
ordinates are truncated below 0.08 c s$^{-1}$.
(c) The same as panel (b), but for TR-2.
   }
 \label{fig:timediv}
 \vspace*{-4mm}
  \end{figure}

\subsubsection{Time evolution of the  modulation effects}
\label{subsec:timediv}
It is known that
the pulse profiles of 4U~0142+61 \citep{Enoto11, Archibold17}
and of some other magnetars (e.g.,\cite{Kuiper12,CotiZelati17})
change on a time scale of a day or longer.
To see whether the demodulation  periodogram changed
across the 3 days covered by the present observation,
we divided the gross time span of 252 ks into two time ranges,
the first 1/3 (about 80 ks in gross after removing the first 6 ks)
to be called Time Range 1, hereafter TR-1,
and the latter 2/3 (about 172 ks) to be called Time Range 2 or TR-2.
Thus, TR-1 and TR-2 cover the 1st day and the latter 2 days, respectively.
This particular time division is justified later.

In figure~\ref{fig:timediv}a,
the solid lines in red and blue give the $N=3$ demodulation periodograms 
derived from TR-1 and TR-2, respectively.
They were calculated over $T=10- 100$ ks,
because of the  shorter data span.
Surprisingly, Zp55 seen  in  figure~\ref{fig:demodulation}a
appears only in TR-2 (blue solid line),
whereas  Zp27 mainly in TR-1 (red solid line).
The parameters of these humps are summarized in table~\ref{tbl:Zp55Zp27}.
We ignore the structures at $\sim 12$ ks, 
which could be instrumental 
because they are close to twice the orbital period (5.8 ks) of NuSTAR.

\begin{table*}[tbh]
\caption{Parameters of the 55 ks and 27 ks humps in the 10--70 keV $Z_N^2$ periodogrms. }
\label{tbl:Zp55Zp27}
\begin{footnotesize}
\begin{tabular}{lccccccccccccc}
\hline 
Time     &      &               &  \multicolumn{5}{c}{Zp55 (= Hump at $T\sim 55$ ks)}     &  
                                         & \multicolumn{5}{c}{Zp27 (=Hump at $T\sim 27$ ks )} \\
                                      \cline{4-8}                                         \cline{10-14}
range   &$N^{*}$&$(Z^2)_0$ &$T$ (ks) &   $Z^2$  & $\Delta Z^2$ & $A$ (s)& $\psi_0$  (deg) &
                                            &$T$ (ks) &   $Z^2$  & $\Delta Z^2$ & $A$ (s) & $\psi_0$ (deg)\\
 \hline \hline 
            & 3 &    465.94&  $54.8 \pm 5.3$  & 476.83 &10.89 & $0.17\pm0.08$   &$60\pm 30$&
                                        &  $26.2\pm 1.7$ & 469.55 & 3.61  & $0.10 \pm 0.05$& $330\pm60$\\
     all   & 7 &    474.50&  $54.3 \pm 5.1$ & 483.67 &9.17 &  $0.15\pm0.09$   &$50\pm 30$&
                                       &  $25.4\pm 2.1$ & 480.28 & 5.78 & $0.14\pm 0.07$  &$340\pm60$\\
 %
           & 3m &  398.18&  $56.6 \pm 4.6$  & 415.96 &17.78 & $0.24\pm0.09$   &$50\pm 30$&
                                       &  $26.4\pm 2.5$ & 406.55  & 8.04  & $0.14 \pm 0.05$ & $300\pm60$\\   
%
  \hline
           & 3  &  143.40 &   ---         &---  &---&---&---& 
                                        &  $26.4\pm 3.1$ &  148.01 & 4.61   & $0.20 \pm 0.12$ &$300\pm 40$ \\                 
TR-1    & 7 &  149.36  & ---            &---   &---&---&---& 
                                        & $25.4 \pm 2.6 $ &   173.98 & 24.62&  $0.39 \pm  0.17$ &$310\pm 30$\\               
              & 7m& 137.41 & ---            &---   &---&---&---& 
                                        & $26.4 \pm 2.9 $ &  166.36 & 28.95&  $0.38 \pm  0.18$ &$310\pm 30$\\               
\hline 
             & 3 &    331.45 &$54.5 \pm 6.2$   &  343.91 & 12.46&   $0.20 \pm  0.11$ &$60\pm 50$ &
                                          &  ---      &---    &---     &---  &---  \\
TR-2    &7  &    341.40 &$54.4 \pm 7.6 $  & 352.76 & 11.36&   $0.22 \pm 0.11$ &$80\pm 50$&
                                          &   ---       &---    &---    &---   &---   \\
             & 3m& 282.82 &$55.0 \pm 7.3$& 299.85 & 17.03&    $0.24 \pm  0.10$ &$50\pm 60$ &
                                         &  ---      &---    &---     &---  &---  \\
 \hline \hline 
\end{tabular}
 $^*$: The $Z^2$ hadmonic number, with ``m" meaning that the secondary pulse peak is masked out
  as described in section \ref{subsec:TwoPeaks}.
\end{footnotesize}
\end{table*}

\begin{figure*}[htb]
  \begin{center}
\includegraphics[height=9.5cm,]{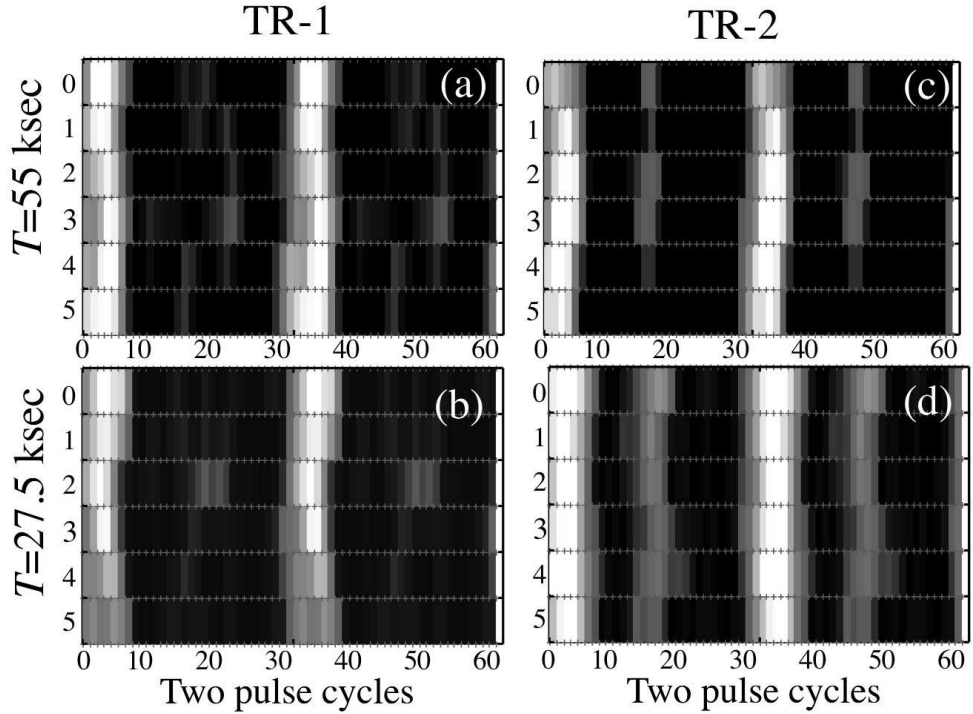}
 \end{center}
  \caption{The same double-folding results of the 10--70 keV photons as figure~\ref{fig:visual}a,
  but derived from TR-1 (panels a and b) and TR-2 (c and d).
  Panels (a) and (c) employ the folding period of $T=55$ ks
  in the ordinate direction,
  whereas (b) and (d) use $T=27.5$ s.}
 \label{fig:55_27.5}
  \end{figure*}

Given the possible $N$-dependence of Zp27
(section \ref{subsubsec:demod_results}),
we repeated the same time-divided analysis by changing $N$.
Then, as shown in figure~\ref{fig:timediv}a by a dashed orange line,
Zp27 in TR-1 became drastically stronger,
until it saturated at $\Delta Z^2 =24.62$ with $N=7$.
As given in table~\ref{tbl:Zp55Zp27}, the amplitude also became 
considerably larger than in equation (\ref{eq:demodpara}).
In contrast, as shown by a dashed cyan line in figure~\ref{fig:timediv}a,
Zp55 in TR-2 did not change significantly  over $N=3-7$.
The widths of these $Z^2$ humps generally satisfy equation (\ref{eq:DeltaT}).
For reference, table~\ref{tbl:Zp55Zp27} also lists the Zp55 and Zp27 information
when the $N=7$ demodulation is applied to the total 252 ks of data.

The different $N$-dependences in TR-1 and TR-2,
as found above rather unexpectedly, suggest
that the 10--70 keV pulse profile in TR-2 is smoother,
whereas that in TR-1 is more structured,
with rich higher harmonic structures
that have been smeared out through the pulse-phase modulation.
To examine this inference, we present in 
panels (b) and (c) of figure~\ref{fig:timediv}
the folded pulse profiles derived separately from the two time ranges.
The raw profile from TR-2 (black in panel c) is similar to 
the time-averaged one in figure~\ref{fig:PGPr_DemNodem},
and changes only slightly (blue) after the demodulation
using the Zp55 parameters given in table~\ref{tbl:Zp55Zp27} (row with TR-2 and $N=3$).
Here and hereafter, the demodulation procedure employs 
$T=55.0$ ks for Zp55 and $T=27.5$ ks for Zp27as fiducial numbers,
considering the latter as half the former.
In contrast, the pulse profile from TR-1 (black in panel b) 
has a more square-wave like shape,
and its edges become sharper (the red profile)
via demodulation using the Zp27 parameters 
in table~\ref{tbl:Zp55Zp27} (row with TR-1 and $N=7$).
These results agree with  the expectation 
from figure~\ref{fig:timediv}a.

The time-evolution effects are further visualized in figure~\ref{fig:55_27.5},
where the  double-folding maps, like figure~\ref{fig:visual}(a),
are presented separately
for TR-1 (left two panels) and TR-2 (right two ones).
The TR-2 data, when folded  at $T=55$ ks (panel c),
approximately reproduce the phase modulation 
seen in  figure~\ref{fig:visual}a,
but such  effects are not observed 
when folded at $T=27.5$ ks (panel d)
at least in the primary pulse.
The TR-1 data, in contrast, 
shows a clear phase  modulation at $T=27.5$ ks (panel b),
which appears also as a hint of two-cycle modulation 
in panel (a) where $T=55$ ks is employed.
These results reconfirm figure~\ref{fig:timediv}.

The phase-modulation properties
thus changed rather drastically from TR-1 to TR-2,
accompanied by some changes in the pulse profile.
Nevertheless, a comparison between
 panels (b) and (c) 
 of figure~\ref{fig:timediv} 
reveals no significant difference
in the average 10--70 keV intensity,
or the pulse fraction.

\begin{figure*}[htb]
  \begin{center}
\includegraphics[width=5.2cm]{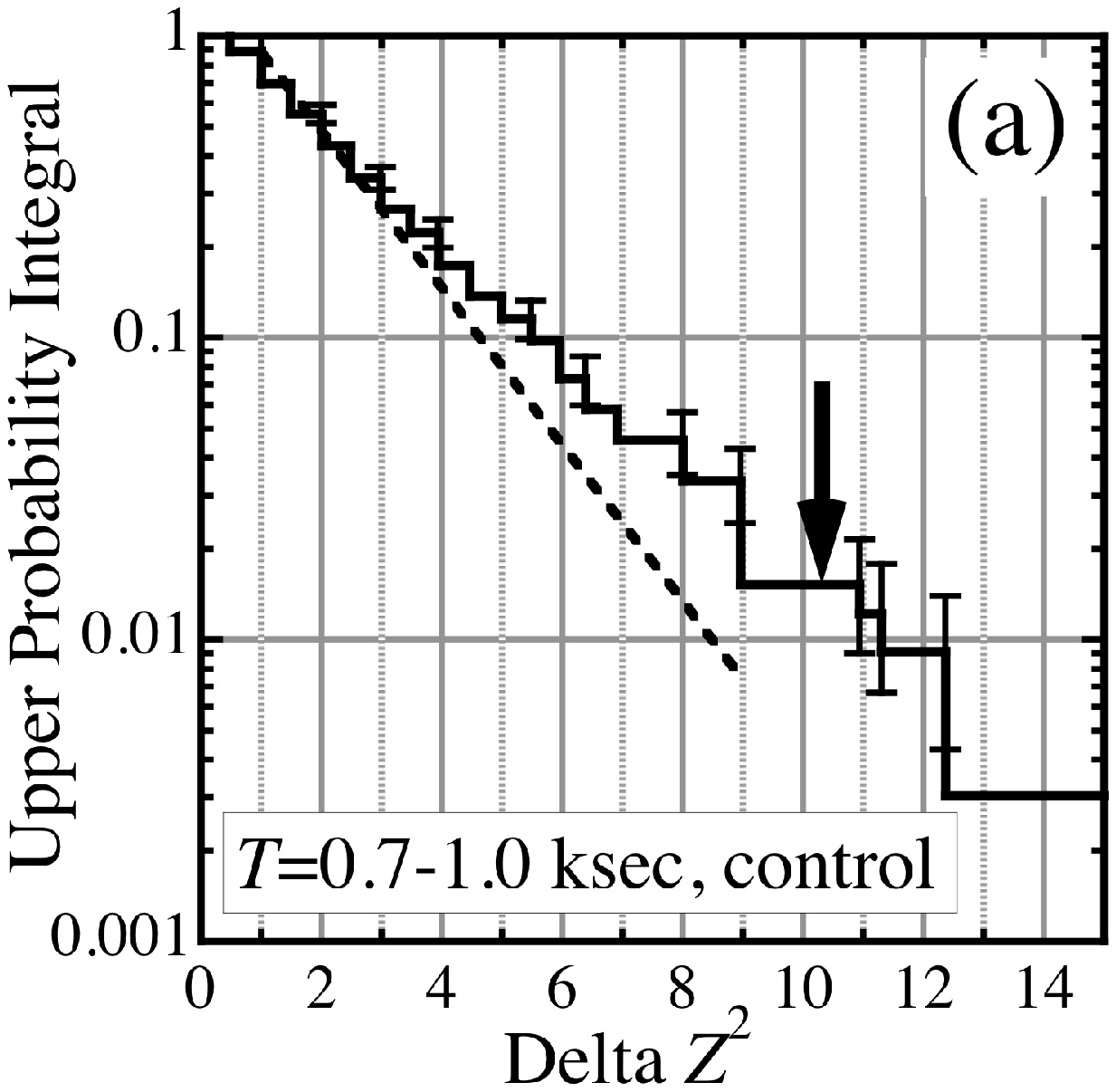}
\hspace{1mm}
\includegraphics[width=5.2cm]{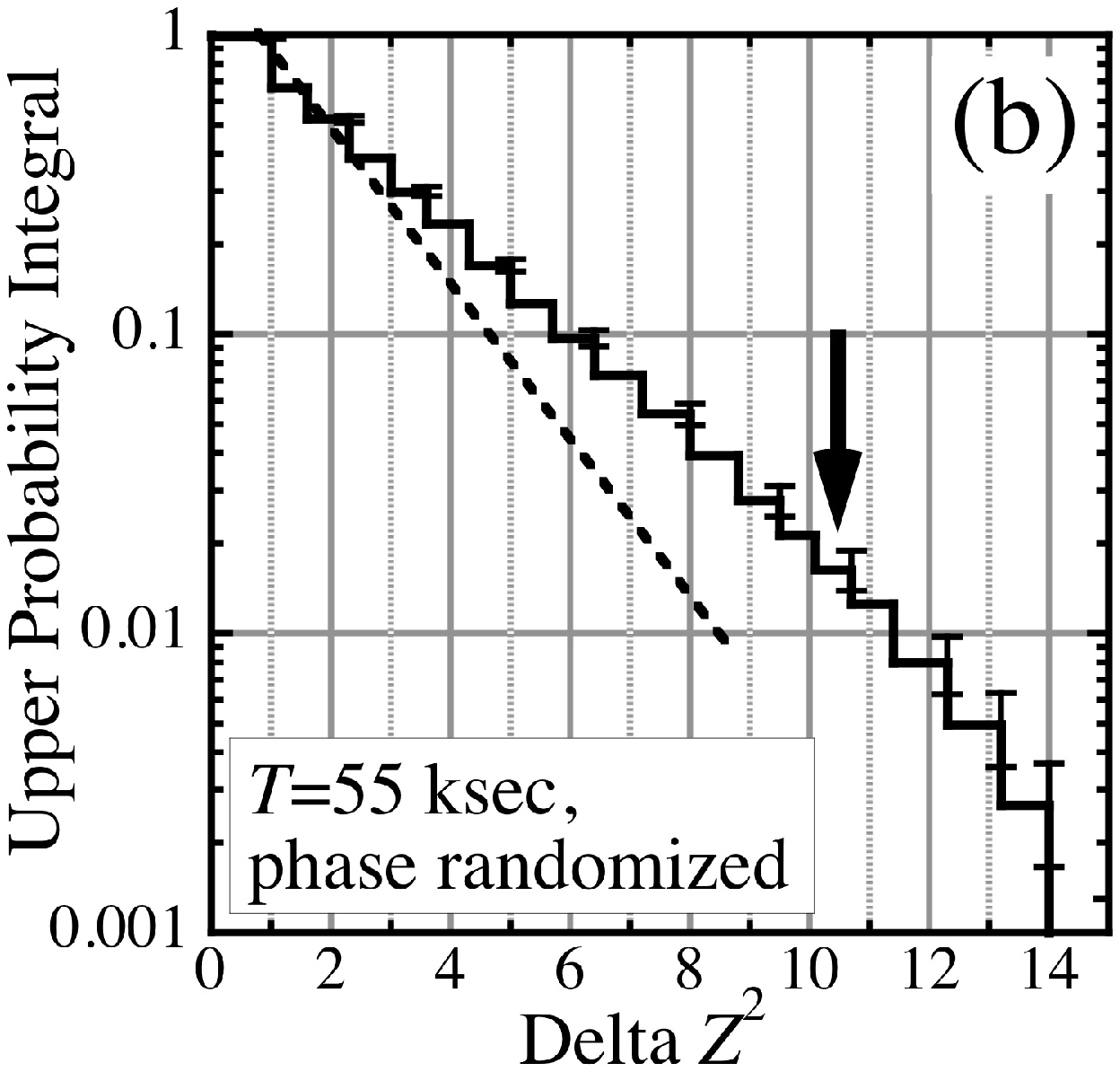}
\hspace{1mm}
\includegraphics[width=5.2cm]{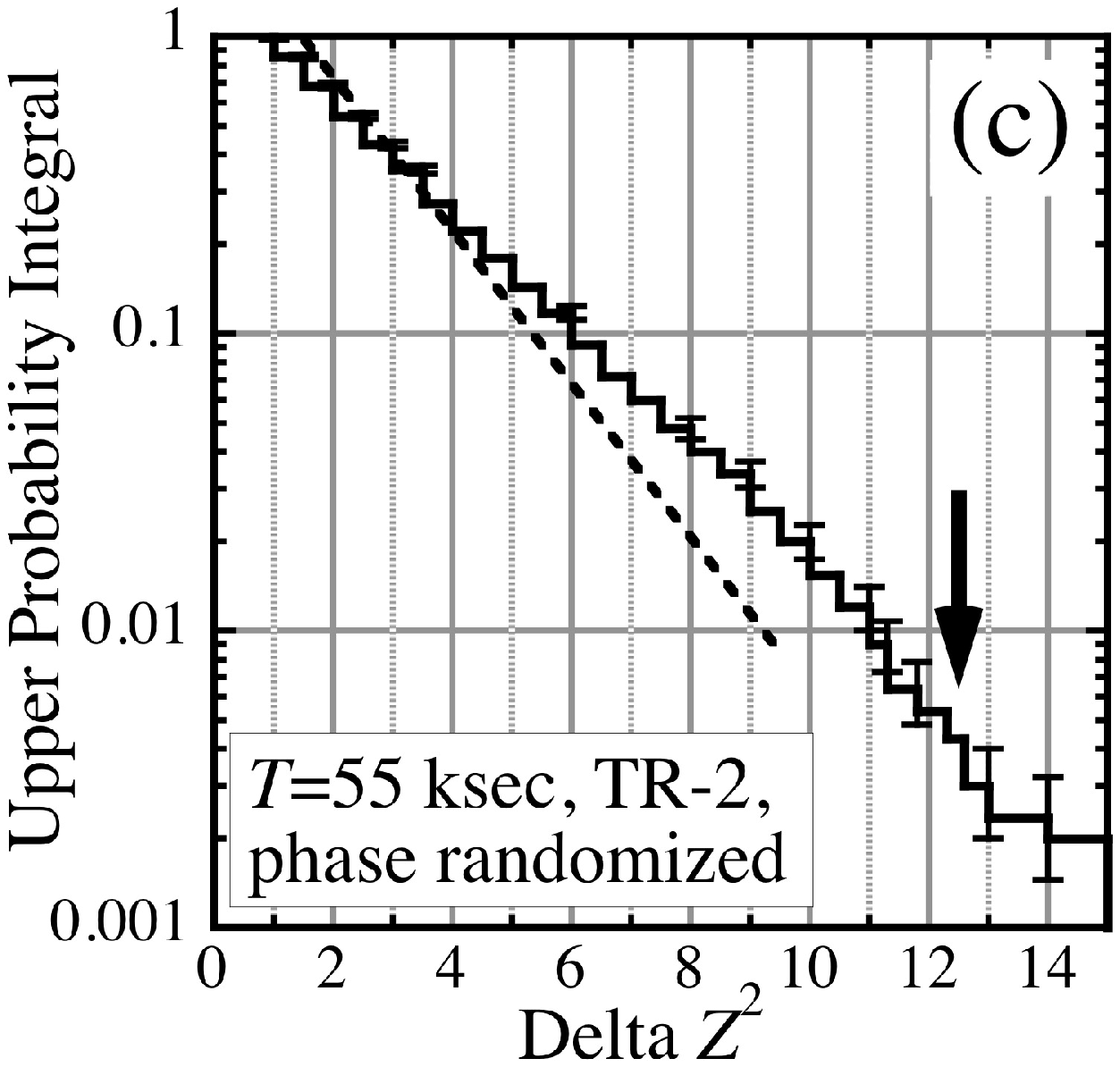}
 \end{center}
  \caption{Estimated upper probability integrals  
  for  $\Delta Z^2$ of equation (\ref{eq:deltaZ2}) in 10--70 keV.
  The dashed line indicates the slope expected for a chi-square distribution
  at $\Delta Z^2 \ll (Z^2)_0$,
  and the vertical arrow shows $ \Delta Z^2$ given in table~\ref{tbl:Zp55Zp27}.
  (a) An estimation based on 300 independent  control trials over $T=0.7-5$ s.
  (b) That based on 3,000 independent trials, all employing $T=55$ ks,
  and incorporating the phase randomization (see text).
   (c) The same as panel (b), but for Zp55 in the 
  $N=3$ demodulation  periodogram from TR-2.
   
  }
 \label{fig:random}
  \end{figure*}

\subsection{Significance of the phase modulation}
\label{subsec:significance}

Although we have so far obtained evidence
that the 10--70 keV pulses are phase-modulated at 55 ks,
or just half that value,
the increments $ \Delta Z^2$ in table~\ref{tbl:Zp55Zp27}
could be due simply to statistical fluctuations,
because $Z^2$ would fluctuate through equation (\ref{eq:Z2_demod}) 
via  different superpositions of the Poisson noises on all  pixels of $C(j,k)$.
To examine this issue, we may adopt a null hypothesis
that the pulse phase is intrinsically {\em not} modulated  
at a given period $T$ beyond Poissonian fluctuations,
and estimate the statistical chance probability ${\cal P}_{\rm dem}$
for such noise superpositions to make  $\Delta Z^2$  at that $T$
higher than a specified value.

Since our aim is to confirm the Suzaku results
rather than to discover a new modulation period
({\it cf.} section~\ref{subsubsec:hardness_ratios}),
the estimation of ${\cal P}_{\rm dem}$  can be made for a single value of $T$, 
neglecting the {\it look-elsewhere} effects,
unlike the cases of Paper I and Paper II.
This is because the range of $T$ in equation (\ref{eq:demodpara})
is  represented by a single wavenumber as
$q=4.6 \pm 0.4$ (for $T_{\rm tot}= 252$ ks),
and hence is effectively covered  by a single independent 
value of $T$ in reference to equation (\ref{eq:DeltaT}).
With this in mind, 
below we estimate ${\cal P}_{\rm dem}$ for Zp55, 
in two different ways both using the actual data themselves.
Monte-Carlo simulations are not employed,
because we encounter difficulties in properly modeling
the intrinsic intensity variations (figure~\ref{fig:LC_5ks})
and the known pulse-profile changes \citep{Enoto11}.

As the first attempt, 
we performed the $N=3$ demodulation analysis,
over a  range of $T=0.7-5.0$ ks,
which is much longer than $P_0$
but shorter than the NuSTAR's orbital period, 5.8 ks.
This  {\it control} study utilized the entire data length (but the first 6 ks), 
and the same scan procedures in $P$,  $A$, and $\psi_0$  as in figure~\ref{fig:demodulation}.
The scan steps in $T$ were chosen to follow equation (\ref{eq:DeltaT}),
to make the results with different $T$ mutually independent.
We thus conducted 300 steps in $T$,
and confirmed that the $\Delta Z^2$ values
from adjacent sampling points of $T$ are uncorrelated.
As a result, $\Delta Z_3^2$ exceeded the target value of 
$\Delta Z_3^2=10.89$ (table~\ref{tbl:Zp55Zp27})
in four realizations among the 300 trials;
this yields ${\cal P}_{\rm dem} \sim 1.3\%$.
More quantitatively, figure~\ref{fig:random}a
shows upper probability integral of the 300 values,
derived by integrating their distribution 
from the largest towards  smaller values.
The probability at first decreases 
as $\propto \exp \left(-\Delta Z^2/2.0 \right)$ (the dashed line), 
as expected for a chi-square distribution of any degree of freedom (Paper II)
at $\Delta Z^2 \ll (Z^2)_0$,
and then it starts exhibiting some excess  at $\Delta Z^2 > 3$,
presumably reflecting the source variability.
This plot yields ${\cal P}_{\rm dem} \sim 2\%$,
which agrees with the above simple-minded calculation.

As the second attempt, we followed Paper II,
and repeated the $N=3$ demodulation analysis of the 252 ks data,
employing the same $(P, A, \psi_0)$ volume and the same parameter steps,
but  fixing the value  $T=55$ ks ignoring the look-elsewhere effect.
Furthermore, in each trial, 
we performed a random permutation as $\{j = 0, 1, .., J-1\}  \rightarrow \{j' = 0, 1, .., J-1\}$,
and replaced  $\{C(j,k) \}$ with $\{C(j',k) \}$
(i.e., a random permutation of rows in figure~\ref{fig:visual}a),
before equation (\ref{eq:Z2_demod}) and equation (\ref{eq:Z2_wiggle}) are executed.
Different trials employed different realizations of the permutation, all with $J=90$.
This phase randomization would  bring any possible intrinsic phase modulation 
at  $T$ to a level that is consistent with Poissonian fluctuations.
Furthermore, all these permutations must be statistically equivalent,
because the Poisson noise in different pixels of $C(j,k)$  should be  independent (Paper II).
As a result, we can study the behavior of $Z^2$ under
the null hypothesis presented above.
We have thus executed 3,000 trials, and found 42 cases 
in which $\Delta Z_3^2$ exceeded $\Delta Z_3^2=10.89$.
More quantitatively, figure~\ref{fig:random}b shows the upper probability integral,
thus derived from the 3,000 trials.
It indicates ${\cal P}_{\rm dem} \sim 2\%$,
in a good agreement with the first evaluation.
We also confirmed that ${\cal P}_{\rm dem}$ is  unchanged
even employing different but similar values of $T$ (e.g., 50 ks).

The latter of the above two evaluations was  applied also to TR-2, 
again with $T$=55 ks fixed and $N=3$.
The results, shown in figure~\ref{fig:random}c,
imply ${\cal P}_{\rm dem} \sim 0.5\%$
for Zp55 in this time range
[blue curve in figure~\ref{fig:timediv}a].
The more stringent value of ${\cal P}_{\rm dem}$ obtained here
is mainly due to the higher target value of  $Z_3^2=12.46$,
because  the upper probability curve is more or less 
similar to that in panel (b).

Finally, the same randomization analysis,
but with $N=7$, was  applied  to the TR-1 data,
to evaluate  significance of Zp27 seen in figure~\ref{fig:timediv}a.
Since the modulation period in this case is less well known in advance,
we scanned over $T$, from 25.0 to 28.5 ks with a step of 0.5 ks.
Then, only in four out of the 3000 trials,
$\Delta Z_7^2$ exceeded the target value of 24.62. 
We hence obtain ${\cal P}_{\rm dem} \sim 0.1\%$,
which is rather conservative (due to the $T$ scan)
but still sufficiently small because of  the very prominent Zp27
in the $N=7$ demodulation periodogram from TR-1.

In summary, Zp55 has a chance probability of 
${\cal P}_{\rm dem} \sim 2\%$ if  the entire data are used,
and ${\cal P}_{\rm dem} \sim 0.5\%$ if  TR-2 is concerned.
Furthermore, Zp27 in TR-1 has a considerably lower chance probability 
as  ${\cal P}_{\rm dem} \sim 0.1\%$.
All these values assume that the modulation period 
is known in advance from the Suzaku result.

\subsection{Energy-dependent effects}
\label{subsec:E-dependence}

\subsubsection{Data in the 10--17.5  and 17.5--70 keV bands}
\label{subsubsec:M&H}

To examine  the phase-modulation phenomenon
for  possible energy dependence,
we subdivided the 10--70 keV band into  two bands,
one at 10--18 keV (called M') and the other at 18--70 keV (called H').
Choosing the boundary at 18 keV, instead of  20 keV, 
is simply to make the pulse significance in terms of $Z^2$ 
about  the same between M' and H';
the results described below do not change
very much  if using the original M and H bands.

\begin{figure}[thb]
  \begin{center}
\includegraphics[width=8cm]{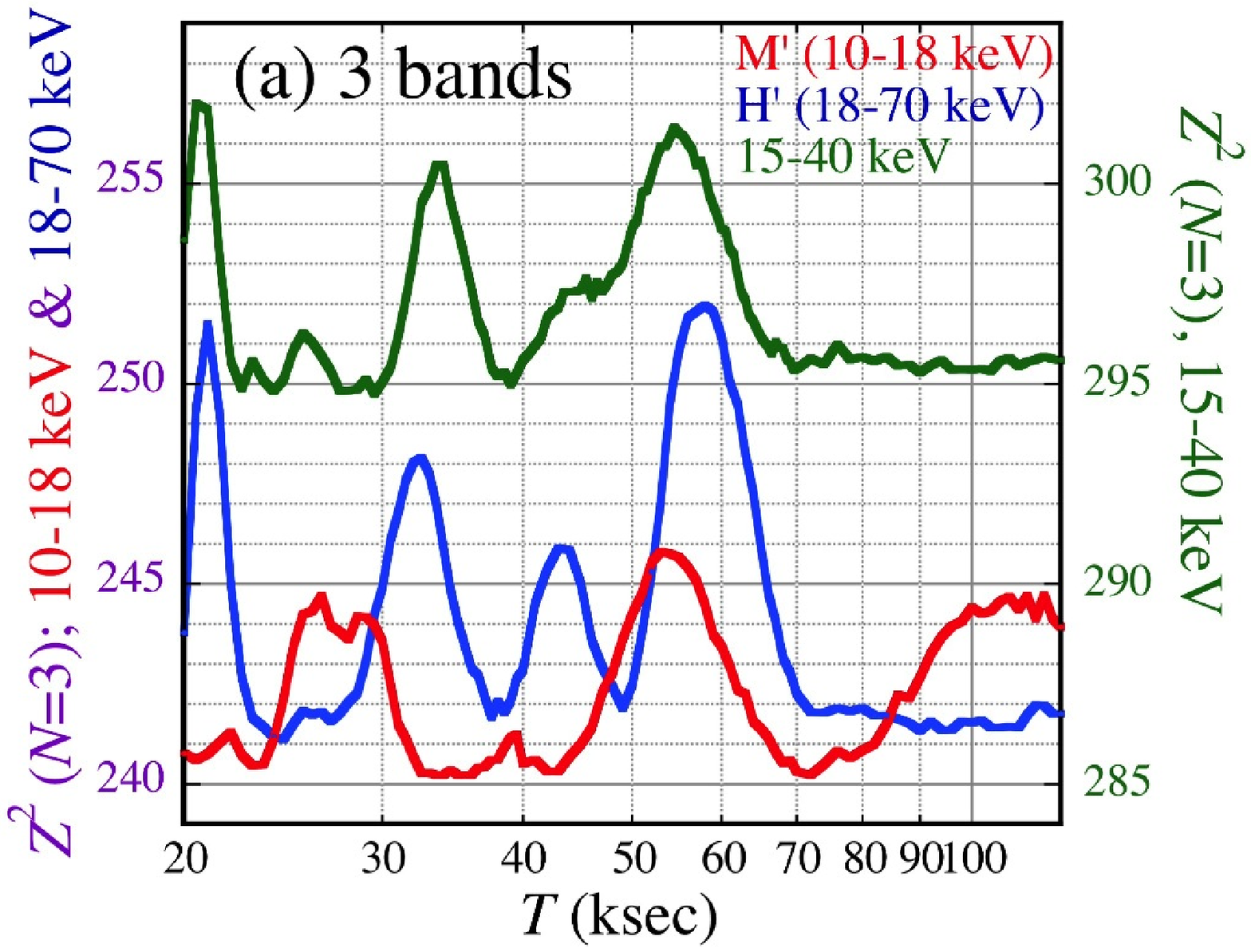}
\includegraphics[width=8cm]{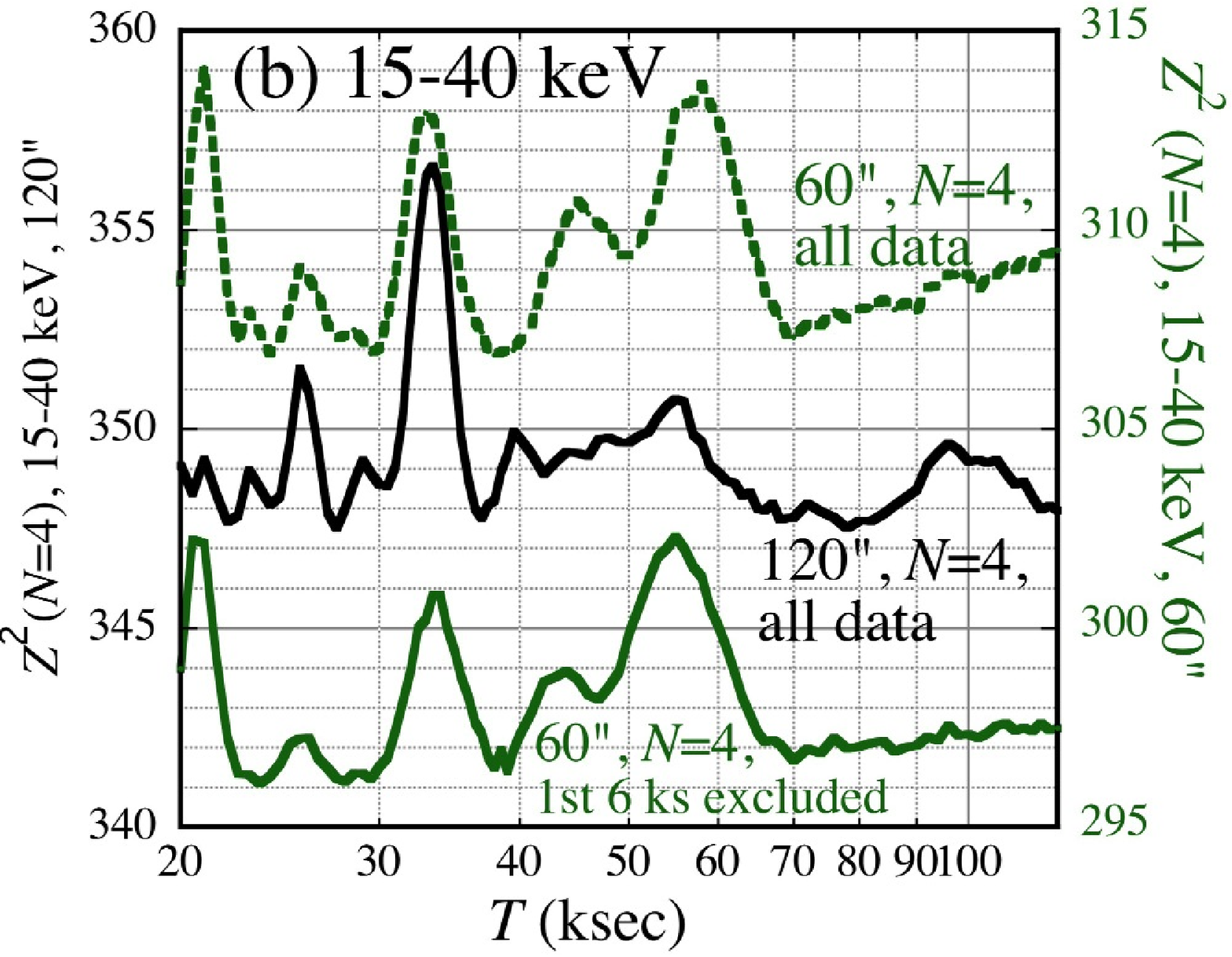}
 \end{center}
  \caption{
(a) Demodulation  periodograms  obtained from 
10-18 keV (M'; red, abscissa to the left), 
18-70 keV (H';  blue, abscissa also to the left),
and 15--40 keV (green, abscissa to the right), 
all using the same analysis conditions as in  figure~\ref{fig:demodulation}a.
(b) Detailed comparison with TEA15, in the 15--40 keV range.
The solid green curve, abscissa to the right,
 is the same as that in panel (a),
but using $N=4$ instead of $N=3$.
The result changes as in the dotted green curve,
when the first 6 ks is restored.
Black (abscissa to the left) shows the $N=4$ result in 15--40 keV 
using the event accumulation radius of $120''$,
to reproduce the condition of TEA15.
  }
 \label{fig:energy_divided}
 \end{figure}

The demodulation analysis in  the M' and H' band,
using the 252 ks data length and $N=3$,
yielded the results presented in 
figure~\ref{fig:energy_divided}a in red and blue, respectively.
Thus, Zp55 is  seen in both demodulation periodograms, 
and is stronger in H'.
Within errors,
the hump centers in M' and H' are both consistent
with equation (\ref{eq:demodpara}).
In contrast, Zp27 is seen only in M'.
Instead, the H' band data exhibits some other enhancements,
at 22 ks, 33 ks, and 42 ks.
Among them, 
the 22 ks feature is likely to be caused by background variations,
because it appears just at 4 times the NuSTAR's orbital period,
and is much stronger in H' than in M'.
The 33  peak can be interpreted as a beat 
between Zp55 in the data, and the one-day periodicity 
which dominates the data-sampling window;
i.e., $1/55 + 1/86.4 = 1/33.6$.
Finally,  the one at 42 ks is  probably a beat 
between Zp55  and two days (172.8 ks).
The 33 ks and 42 ks features are also stronger in H'  than in M',
probably because these beat effects can be enhanced
when  Poisson fluctuations of comparable strengths
are added in quadrature to the intrinsic beat signal.

\subsubsection{Consistency with Tendlukar et al. (2015)}
\label{subsubsec:TEA}
Analyzing the same NuSTAR data 
over the 15--40 keV range (the same conditions as in Paper I),
TEA15 found no evidence for the pulse-phase modulation 
at any period between 45 ks and 65 ks,
including in particular 55 ks.
To cross check  their report,
we applied the demodulation analysis to the 15--40 keV data,
using the same analysis conditions and the same parameter search steps
(including 0.01 s in $A$) as in figure~\ref{fig:demodulation}.
The result  is shown in green in figure~\ref{fig:energy_divided}a.
There, we still find Zp55, though with a reduced heights
due presumably to the narrower energy range.

Although the above finding might appear inconsistent with TEA15,
our analysis conditions still differ from those of TEA15 in several points.
In fact, TEA15 used $N=4$ instead of our $N=3$,
and they did not discard the 1st 6 ks.
More importantly,
they used the event accumulation region of $120''$ radius  (vs. ours of $60''$)
so that their data have larger  background contributions,
which are estimated to be 16\% in M and 64\% in H.
To  progressively better reproduce the conditions employed by TEA15,
we first changed from $N=3$ to  $N=4$, and obtained 
the  solid green curve in figure~\ref{fig:energy_divided}b.
Its  behavior is very similar to the case of $N=3$ in panel (a),
except a systematic increase in $Z^2$ by 1.5-2 which is reasonable.
Next, we restored the first 6 ks of the data,
and obtained the dotted green periodogram in panel (b).
Thus, the inclusion/exclusion of the 1st 6 ks
 systematically changes the $Z^2_3$ values by $\sim 12$,
 but no other significant differences are noticed. 
 Finally, we extracted FPMA+FPMB events within $120''$ of the image centroid,
instead of $60''$,
and applied the demodulation analysis to the 15--40 keV data with  $N=4$.
Then, as  shown in black in figure~\ref{fig:energy_divided}b, 
the $Z^2$ values generally increased by $35-40$
due to the increased signal photons,
but Zp55 became no longer visible,
probably buried by increased statistical fluctuations.
Instead, the beat at 33 ks (which was outside the range searched by TEA15)
increased significantly, 
presumably again due to a coupling between 
the intrinsic 55 ks periodicity and  larger background fluctuations.
Under such conditions, 
a beat could sometimes dominate over the original signal.

We have thus reproduced the essential point by TEA15,
and confirmed the consistency between the two studies.
The overall comparison can be interpreted in the following way.
In the Suzaku observation,
the signal-to-noise ratio of the demodulation periodogram 
was optimized by selecting a relatively limited 
energy band (15--40 keV) to reduce background,
because the HXD is a non-imaging device,
and yet the modulation amplitude was rather large 
($A \sim 0.7$ s; Paper I) on that occasion.
In the present case, in contrast,
the difficulty in confirming the 55 ks modulation 
lies  primary  in the very small value of $A$.
In addition, the sensitivity with NuSTAR, 
being an imaging instrument, would be optimized
by widening the signal-acceptance energy band,
and by choosing a narrow event accumulation region
to suppress the background.

\begin{figure}[htb]
  \begin{center}
\includegraphics[width=8.cm,]{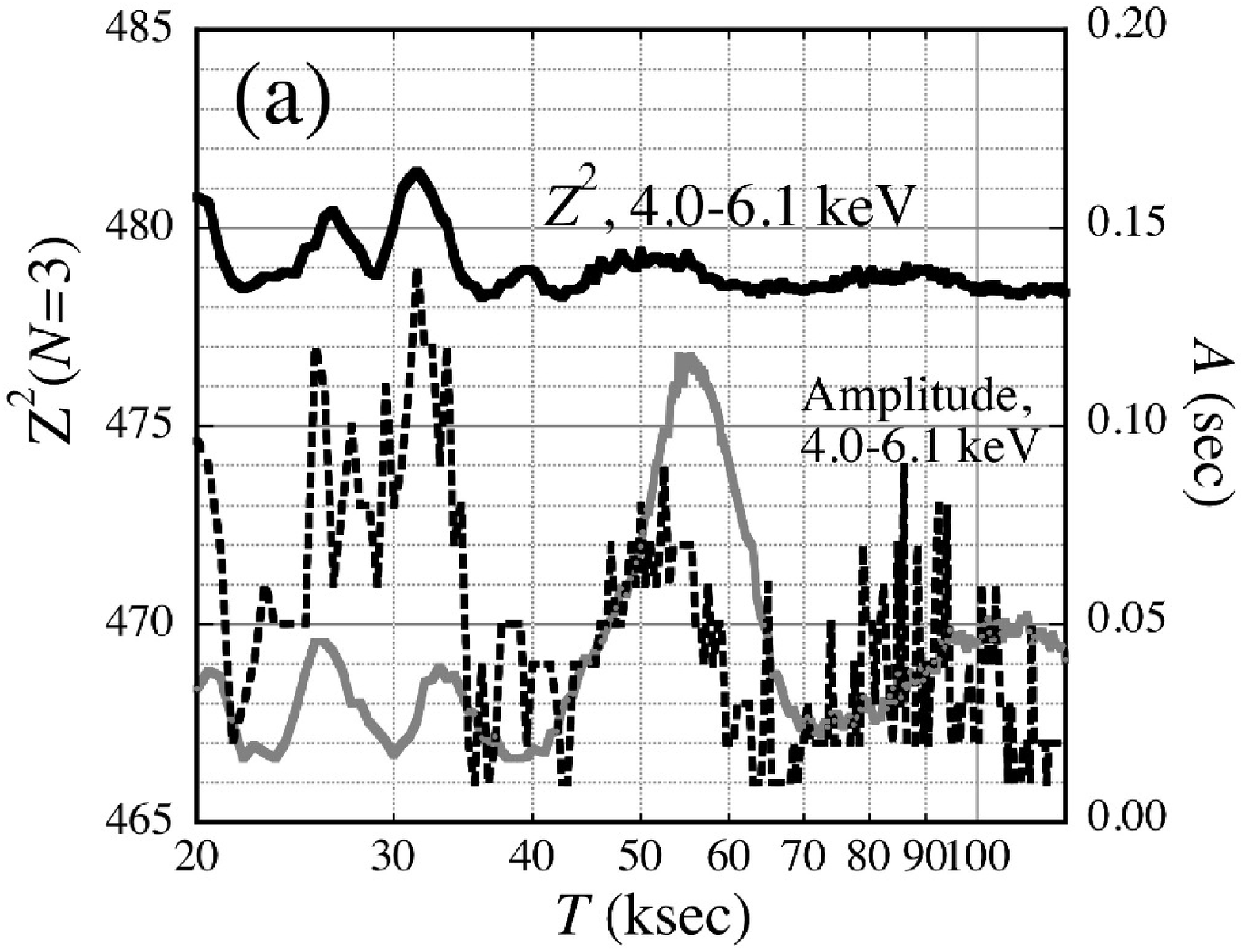}
\hspace*{5mm}
\includegraphics[height=5.0cm,]{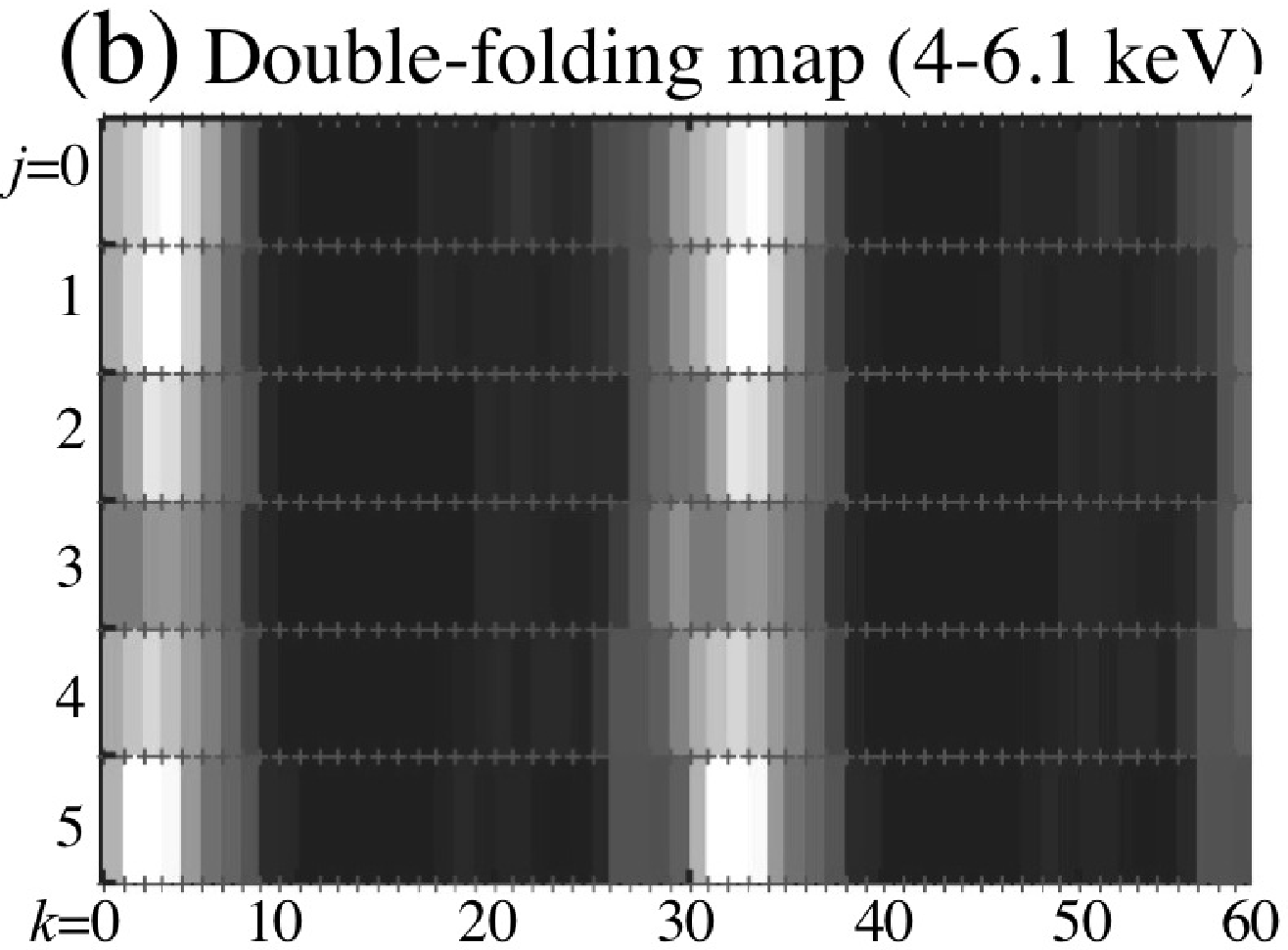}
\includegraphics[height=6.5cm,]{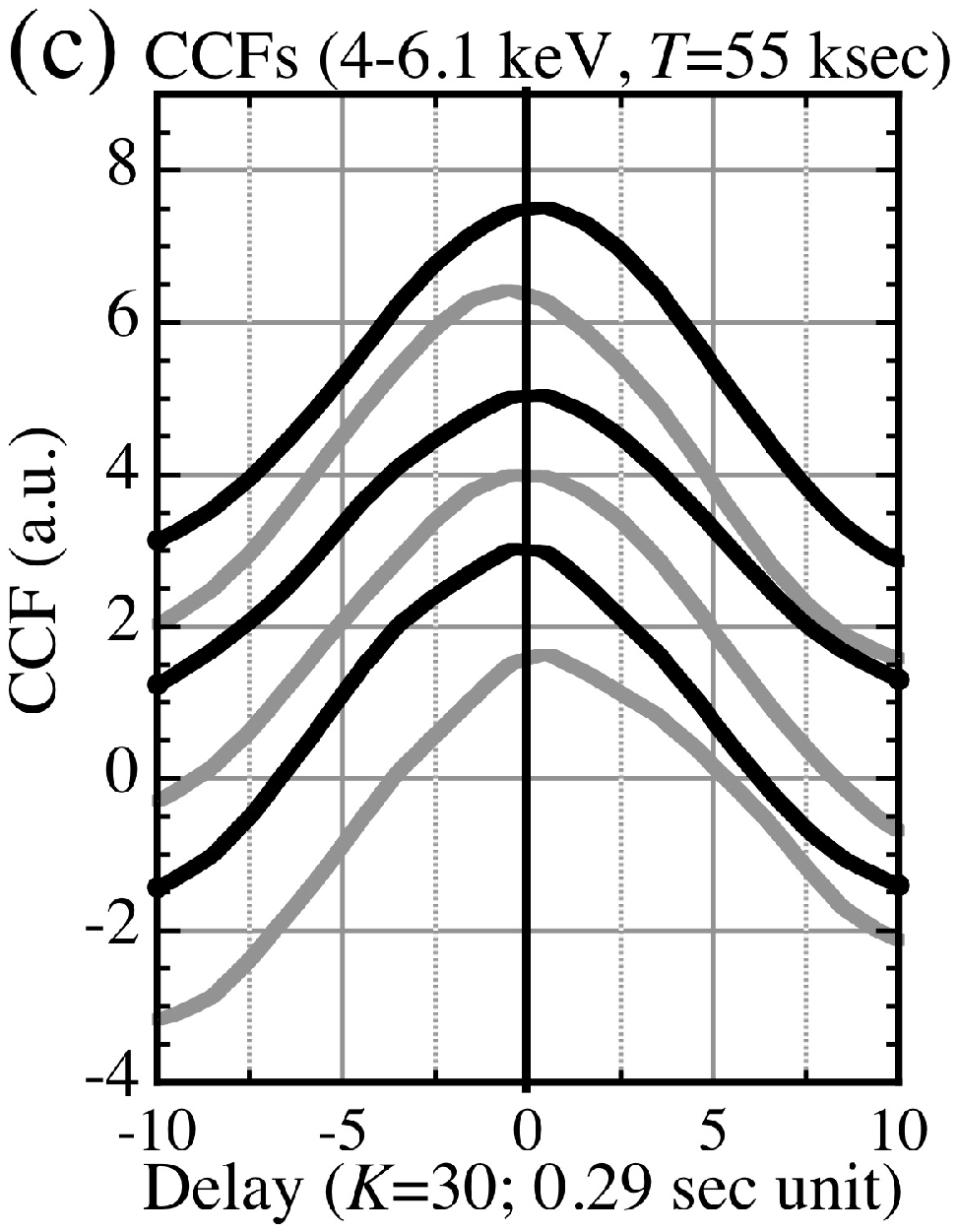}
 \end{center}
  \caption{
 Results obtained from low-energy bands.
 (a) The $N=3$ demodulation periodogram in 4--6.1 keV (solid curve),
 and the behavior of $A$ (dashed curve). 
 The  10--70 keV result, 
 the same as in figure~\ref{fig:demodulation}a,
 is superposed in gray.
 (b)  The same double-folded map as figure~\ref{fig:visual}a, 
 but for the 4--6.1 keV signals.
 (c) The same CCFs as figure~\ref{fig:visual}c,
 derived in the 4--6.1 keV range.
   }
 \label{fig:demodSXC}
  \end{figure}

\subsubsection{Analysis of  lower-energy data}
\label{subsubsec:lowE}

It is important to examine
whether Zp55 and Zp27 are  also present  in the SXC,
which dominates the energies below $\sim 10$ keV.
We hence  performed the same  demodulation analysis with $N=3$,
using the 4--6.1 keV photons.
This particular energy range was chosen
because  the value of $(Z^2)_0=477.40$  therein
is close to that in the 10--70 keV range,
enabling an easy comparison.
The result,  given in figure~\ref{fig:demodSXC}a,
shows no evidence of significanct $Z^2$ increase at $T\sim 55$ ks,
although some structures are observed at $T\sim 27$ ks and $\sim 31$ ks.
Tentatively, an upper limit of $A<0.1$ s may be assigned,
where 0.1 was taken as an average of errors in $A$ in table~\ref{tbl:Zp55Zp27},
and also considering the behavior of $A$ in 4--6.1 keV at $T \sim 55$ ks.
We obtain  essentially the same result
even when expanding the energy band into 4--10 keV,
where we find  $Z^2 \sim 785$ on average.

Panels (b) and (c) of figure~\ref{fig:demodSXC} 
provide the same information as figure~\ref{fig:visual}a and figure~\ref{fig:visual}c,
respectively, for the 252 ks of 4--6.1 keV signals.
Because the pulse profile is nearly single-peaked in low energies (e.g., TEA15),
a single ridge per cycle  is observed in panel (b).
No clear evidence of  sinusoidal phase modulation
at $T=55$ ks is observed, either in panel (b) or (c),
although the CCFs suggest some higher harmonic effects
as already seen in figure~\ref{fig:demodSXC}a.
This is possibly in line with figure~\ref{fig:energy_divided},
where Zp27 is suggested to be softer than Zp55.

In short,
the result obtained here agrees with the finding made in Paper I 
from 4U~0142+61 (and that in Paper II from 1E~1547$-$5408 as well),
that the phase modulation is seen only in the HXC.

\subsection{The primary and secondary pulse peaks}
\label{subsec:TwoPeaks}
The pulse profiles of 4U~0142+61 above 10 keV are double-peaked,
consisting of the primary  and secondary peaks
(figure~\ref{fig:PGPr_DemNodem}b, figure~\ref{fig:timediv}).
The demodulation results so far obtained for the HXC
are considered to mainly reflect the behavior of the primary peak,
because it is significantly stronger.
To confirm this inference,
we conducted the same analysis,
but  {\it masking out} the secondary peak;
that is, after processing the data with 
equation (\ref{eq:Z2_demod}) and equation (\ref{eq:Z2_wiggle}),
the pulse profile $D(k)$  from $k=10$ to $k=23$ (for $K=30$) 
in figure~\ref{fig:visual}b
(from $120^\circ$ to $270^\circ$ in the pulse phase)
is replaced by the pulse-phase average of $D(k)$,
and then the Fourier transform is executed to calculate $Z^2$.
As listed again in table~\ref{tbl:Zp55Zp27}, 
this analysis has given results that are rather  similar to
figure~\ref{fig:demodulation} and figure~\ref{fig:timediv}.
We therefore confirm that the results obtained in section~\ref{subsec:demod} 
mainly reflect the behavior of the primary pulse peak.

On a closer look at table~\ref{tbl:Zp55Zp27}, 
both $\Delta Z^2$ and  $A$ of Zp55
are  found to generally increase
by masking out the secondary pulse peak.
Therefore,  the secondary pulse peak is 
likely to be free from the phase-modulation effects,
and is partially canceling out the modulation 
specified by the primary peak.
This may explain why the values of $A$ of Zp55,
which is $\sim 0.17$ s from the demodulation periodogram 
in figure~\ref{fig:demodulation}a 
and $\sim 0.15$ s from the CCF in figure~\ref{fig:visual}c,
are smaller than those visually suggested by 
figure~\ref{fig:visual}a and  figure~\ref{fig:visual}b.

For further examination,
we tried to mask out the primary pulse peak,
namely, $k=26$ to $k=40$ (for $K=30$) in figure~\ref{fig:visual}b.
However, the reduced statistics did not allow us 
to obtain meaningful results.

\section{Discussion}
\label{sec:discussion}

\subsection{Summary of the NuSTAR results}
\label{subsec:summary}

Analyzing the NuSTAR data  of 4U~0142+61 acquired in 2014,
we have obtained the following results.

\begin{enumerate}

\item
As in figure~\ref{fig:demodulation},
the HXC pulses  are  phase-modulated, 
at a period  of $\sim 55$ ks [equation~(\ref{eq:demodpara})]
which is consistent with the Suzaku period of equation (\ref{eq:55ks}).
When only the range of equation (\ref{eq:55ks}) is considered employing $N=3$,
this peak, Zp55, is significant at 98\% confidence if using the entire NuSTAR data,
and at  $99.5\%$ if the time region is limited to TR-2
(section~\ref{subsec:significance}).

\item
The phase-modulation amplitude associated with Zp55
 is rather small, $A \sim 0.17$ s, or $A/P \sim 0.02$.

\item
The data also exhibit Zp27 (figure~\ref{fig:timediv}a), 
i.e., an indication of HXC pulse-phase modulation 
at a period  that is consistent with just half the value 
of equation~(\ref{eq:demodpara}) and equation~(\ref{eq:55ks}).

\item
From TR-1 to TR-2, neither the average 10--70 keV intensity
nor the pulse fraction changed significantly.
Nevertheless, Zp27 and Zp55 appeared 
essentially only in TR-1 and TR-2, respectively,
involving some changes in the HXC pulse shapes 
(section~\ref{subsec:timediv}).

\item
Particularly in TR-1, the Zp27 significance increased with $N$,
from $\Delta Z_3^2=4.61$ to $\Delta Z_7^2=24.62$,
the latter implying $99.9\%$ confidence.
In contrast, the Zp55 significance depends little on $N$.

\item
The 27 ks periodicity  appears also  
in the power density spectrum of the H/M ratio,
at a wavenumber $q=9$ (figure~\ref{fig:HR_periodicity}).
This $q=9$ peak is significant at a confidence level of 
$\sim 90\%$ or higher,
if the {\it look elsewhere effects} are ignored.

\item
The negative detection of these effects by TEA15,
using the same NuSTAR data,
can be attributed primarily to the rather small value of $A$,
augmented by  the narrower energy band
and the wider data accumulation radius they employed
(section~\ref{subsubsec:TEA}).

\item
While the primary peak of the HXC pulse is  phase modulated,
the secondary peak is probably not (section~\ref{subsec:TwoPeaks}).

\item
Below 10 keV where the SXC dominates,
the pulse phase is modulated
at neither  $\sim 55$ ks nor $\sim 27$ ks,
with a typical upper limit of $A <0.1$ s
(section~\ref{subsubsec:lowE}).

\end{enumerate}

\begin{figure}[htb]
  \begin{center}
\includegraphics[width=8cm,]{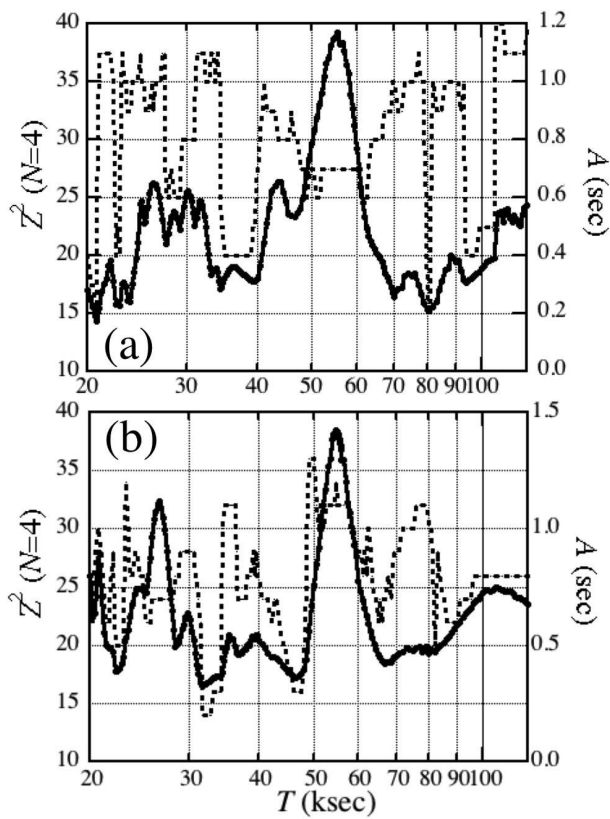}
 \end{center}
  \caption{
Suzaku results on the 15--40 keV pulse-phase modulation,  
 obtained in 2009 (panel a) and 2013 (panel b),
 with the former updating figure 2d of Paper I.
 The solid curves show the $Z_4^2$ perodograms (ordinate to the left),
 and the dashed curves the value of $A$ 
 (ordinate to the right) that maximizes $Z_4^2$ at each $T$.
The associated parameters are given in table~\ref{tbl:4datasets}.
  }
 \label{fig:Suzaku0913}
  \end{figure}

\subsection{Brief (re)-analysis of  the Suzaku data}
\label{subsec:Suzaku0913}
To complement the results from NuSTAR,
we conducted two brief (re)-analyses of the
Suzaku data of 4U~0142+61.

One is a re-analysis of the 2009 Suzau data,
because the $N=4$ demodulation periodogram in figure 2d of Paper I
covered only $30 \leq T\leq 70$ ks,  without information at $\sim 27$ ks.
We hence repeated the same demodulation analysis as in Paper I, 
again using the 15--40 keV HXD data in 2009 with  $N=4$,
but scanning $T$ over a wider range of  20--120 ks 
with a finer step of 0.2--2 ks than in Paper I;
these parameters  match those in the present NuSTAR study.
The other parameters were scanned in the same way as in Paper I,
namely,  $A=0 -1.2$ s with 0.1 s step,
$\psi_0=0^\circ-360^\circ$ with  $10^\circ$ step,
and $P=8.68894-8.688904$ s with 5 $\mu$s step.
The obtained results, shown in figure~\ref{fig:Suzaku0913}a,
reproduce the data points in figure~2d of Paper I,
and reconfirm the 55 ks peak,
of which the parameters are summarized in table~\ref{tbl:4datasets}.
We also notice a hint of peak at $T\sim 27$ ks,
although its significance is not obvious.

The other concerns the Suzaku data of 4U~0142+61 obtained in 2013.
Below, we concisely describe the observation and  the obtained results,
leaving further details to a later publication.
Aiming to reconfirm the discovery described in Paper I,
this follow-up  Suzaku observation was conducted
from 09:46:00 UT on 2013 July 31 to 15:00:00 UT on August 2,
for exposures as given in table~\ref{tbl:4datasets}.
The observing conditions and the source intensity
were very similar to those in 2009.
The pulsation was detected with the XIS
at $P=8.68914 \pm 0.00003$ s,
and the demodulation analysis was  applied
to the background-inclusive 15--40 keV HXD data with $N=4$.
We employed the same parameter search conditions
as in figure~\ref{fig:Suzaku0913}a,
except a wider range of the amplitude scan as $A=0-1.5$ s.
The results are presented in figure~\ref{fig:Suzaku0913}b
in the same manner as figure~\ref{fig:Suzaku0913}a
and figure~\ref{fig:demodulation}a,
and the derived parameters are summarized in table~\ref{tbl:4datasets}.
Thus, the 55 ks peak is clearly confirmed,
with a slightly smaller $\Delta Z^2$
and a possibly larger $A$ than in 2009.
Furthermore, we observe a clear $Z_4^2$ peak at $T\sim 27$ ks.
Like in 2009, the 1--10 keV pulses showed 
no evidence of phase modulation at  55 ks or 27.5 ks,
with upper limits of $A< 0.3$ s 
considering the low time resolution of the XIS (Paper I).

\begin{table*}[hbt]
\caption{Parameters of the 55 ks peak in the demodulation  periodograms from the four data sets.}
\label{tbl:4datasets}
\begin{center}
\begin{tabular}{llcccrcccc}
\hline
\hline
 Data set                               & Exp. (ks)     & Energy &$N$   &Period$^*$ & $(Z_N^2)_0~^\dagger$& $T$    &$\Delta Z_N^2$   & $A$ \\
                                               &  gross/net   & (keV)     &          &     (s)           &                    & (ks)       &      at $T$              &  (s)   \\
\hline
Suzaku 2007 Aug$^\ddagger$                & 204/101& 15--40 &4& 8.68878 & 52.01      &---  $^\S$          & ---  $^\S$    &  ($<0.9$)$^\S$  \\
Suzaku 2009 Aug$^{\ddagger \|}$           &186/102 & 15--40  &4& 8.68899 &13.87     & $55 \pm 4$      &25.63   &$0.7 \pm 0.3$\\
Suzaku   2013 Jul-Aug $^{\|}$                &192/95  & 15--40  & 4& 8.68914 & 16.36    & $54 \pm 3$       &  22.17 & $1.2 \pm 0.4$\\
NuSTAR 2014 March                                &258/168 & 10--70 & 3&  8.68917 &465.94 & $54.8 \pm 5.3$ &10.89 & $0.17 \pm  0.08$\\
\hline
\end{tabular}
\begin{enumerate}
\item[$^*$]: with a typical uncertainty of $\pm (2-5)\times 10^{-5}$ s.
\item[$^\dagger$]: at the peak of the pulse periodogram  incorporating no demodulation.
\item[$^\ddagger$]: from Paper I, but with some minor updates.
\item[$^\S$]: no significant $Z^2$ peak was found at  $T\sim 55 $ ks.
\item[$^\|$]: see figure~\ref{fig:Suzaku0913}.
\end{enumerate}
\end{center}
\end{table*}

\subsection{Comparison with the Suzaku results}
\label{subsec:comparison_Suzaku}
We are now ready to examine the nine items in section~\ref{subsec:summary},
through their comparison with the Suzaku results
given in Paper I and section~\ref{subsec:Suzaku0913}.
First of all, item 1 on Zp55 agrees with 
the Suzaku achievements in 2009 and 2013.
In fact,  the value of $T \sim 55$ is fully consistent
among the three observations (table~\ref{tbl:4datasets}),
and the three demodulation periodograms 
(figure~\ref{fig:demodulation}a, figure~\ref{fig:Suzaku0913}a, 
and figure~\ref{fig:Suzaku0913}b) are strikingly similar,
even though $A$ is different.
Another  common property of the three data sets
is the absence of these effects in the SXC (item 9).
We hence conclude 
that the 55 ks pulse-phase modulation is not a Suzaku-specific artifact,
but is an intrinsic property of 4U~0142+61
that is  also observed with NuSTAR.
The phenomenon is absent (or much weaker) in the SXC.

In  Paper I, 
the chance probability was calculated as  $<1\%$ for the 55 ks peak 
with $\Delta Z_4^2=25.63$ (figure~\ref{fig:Suzaku0913}b, table~\ref{tbl:4datasets})
to appear {\em at any period} over  $T=30-70$ ks in the 2009 Suzaku data.
In the present work, that of finding the 55 ks peak  with $\Delta Z^2=10.89$ 
(table~\ref{tbl:Zp55Zp27}, table~\ref{tbl:4datasets})   
in the NuSTAR data, {\em at a consistent period},
has been estimated as $\lesssim2 \% $ (section \ref{subsec:significance}).
An evaluation similar to that in figure~\ref{fig:random}
yields a chance  probability of $\sim 1\times 10^{-4} $ 
for the 2013 Suzaku data to show the $Z^2$ peak  
with $\Delta Z^2 =22.17$  (table~\ref{tbl:4datasets})
{\em  at the same period} (figure~\ref{fig:Suzaku0913}b).
Multiplying these three numbers,
the overall chance probability to observe  the 55 ks HXC 
pulse-phase modulation in the three observations
can be reduced to $\sim 2 \times 10^{-8}$.

Item 2 provides the largest difference of the present NuSTAR results,
compared with those from Suzaku.
Indeed, the value of $A$ has decreased by 4 and 7 times
from the 2009 and 2013 Suzaku observations, 
respectively, the latter only in less than a year
from 2013 August to  2013 March (the present NuSTAR observation).
The change of $A$, however, 
had already been noticed in Paper I,
because the HXC pulse-phase modulation was absent
in  the Suzaku data acquired in 2007 \citep{Enoto11},
when the observing conditions and the source intensity 
were both similar to those in 2009 and 2013.
The results from the 2007 Suzaku observation are quoted in table~\ref{tbl:4datasets}.

In this regard, the present NuSTAR data 
resemble those of Suzaku in 2007,
in which the HXC pulsation was detected significantly
with $(Z_4^2)_0=52.01$ for $N=4$  (table~\ref{tbl:4datasets}),
through the simple period search \citep{Enoto11}
without considering the pulse-phase variations.
In contrast, the HXC pulsation in the 2009 and 2013 Suzaku data 
was rather insignificant  if employing such a conventional analysis,
with $(Z_4^2)_0=13.87$ (upper probability of 8.5\%) in 2009 
and $(Z_4^2)_0=16.38$ in 2013 (3.7\%).
It was only after the application of the demodulation procedure
that the HXC pulsation  in these data sets were detected significantly,
with $Z^2>38$ (table~\ref{tbl:4datasets}),
at the same period as the SXC pulsation.
We thus conclude that $A$ can vary considerably,
on a time scale of months to years;
sometimes $A$ is so large as to smear out  the HXC pulses (2009 and 2013),
but  on other occasions the effect is undetectable (in 2007)
or detectable only through a very careful analysis 
of high-quality data, just as in the present case.

Another important effect found with NuSTAR is item 3,
i.e., the appearance of Zp27 at about half the period of Zp55.
Although this issue was not considered in Paper I,
the reanalysis in figure~\ref{fig:Suzaku0913}a 
suggests its hint in the 2009 Suzaku data.
Moreover, it is much more clearly seen 
in those of 2013 (figure~\ref{fig:Suzaku0913}b).
The 27 ks periodicity is further supported by item 6,
because  the $q=9$ peak  in the power spectrum of the H/M ratio 
(figure~\ref{fig:HR_periodicity}) has a period  
which is consistent with  $T\sim 27$ ks.
We hence conclude that the periodicity at $\sim 27$ ks 
is another intrinsic property of this magnetar,
and regard Zp27 as the 2nd harmonic (in terms of frequency) of Zp55,
even though the relation between the H/M variation
and the pulse-phase modulation  is not obvious.

Still more intriguing findings are item 4,
i.e., the switching between Zp27 and Zp55,
and item 5, i.e., the strong $N$-dependence of Zp27.
Just as $A$ of Zp55 is variable,
the amplitude of Zp27 would also vary,
either independently of that of Zp55, 
or in a mutually exclusive manner.
Although we searched the 2009 and 2013 Suzaku data for similar effects,
the results were inconclusive due to insufficient statistics.

\subsection{Possible origins of the modulation}
\label{subsec:possible_origins}
Our discussion has so far been purely empirical.
Before conducting more physical interpretations in terms of free precession
of a rigid body (section~\ref{subsec:free_precession}), 
we survey, once agin, other  possible explanations of what has been observed.

As argued in Paper I,
the pulse-phase modulation with  $T=55$ ks and $A=0.7$ s 
(the 2009 Suzaku parameters) could have been 
explained in the most natural way
if the source formed a binary with a companion  
which has a mass $0.12~M_\odot/\sin{i}$,
where $M_\odot$ is the solar mass and $i$ the inclination.
However, this interpretation is clearly ruled out,
by the drastic change in $A$ in 5 months,
and the absence of the same modulation in the SXC.
This conclusion, already derived in Paper I,
has  been much reinforced by the present study.

Supposing that 4U~0142+61 is an isolated NS,
it could still be powered by accretion from 
a fall-back disk (e.g., \cite{fallback_disk}).
Then, a long periodicity could arise via precession of the assumed disk.
As the disk precesses,
the materials may be caught by different magnetic field lines of the NS,
so that the accretion columns (emitting the HXC) could be slightly displaced,
to produce a periodic shift in the pulse phase.
Fluctuations of the disk conditions could change $A$.
If, in addition, the SXC arises from the heated NS surface,
its pulses would be free from phase jitters.
In this way, major  observational results could be explained.
However, such a disk without binary perturbation should precess 
on a time scale longer than a few years \citep{disk_precession},
and the precession period would fluctuate.
Moreover, under such a condition,
the X-ray intensity would vary at the same period.
To test this possibility, we folded the 10--70 keV NuSTAR data
at 55 ks and 27.5 s, into 8 phase bins,
but the folded profiles were quite flat,
with chi-squared of 7.46 and 5.90 for 7 d.o.f., respectively.
This is consistent with the absence of noticeable power
in the red and blue power spectra in  figure~\ref{fig:HR_periodicity}
at $q  \sim 9$ or $q \sim 4.5$.
Thus,  the disk precession view is disfavored.
A similar study using the Suzaku data was  inconclusive,
due to the higher background.

Evidently,  magnetars are more likely to be powered by
their magnetic energies \citep{Magnetar,Mereghetti08,Enoto10},
which are somehow released into radiation
at regions probably near their magnetic poles
as evidenced by their pulsed X-ray emission.
Since the pulse profiles of 4U~0142+61 are variable \citep{Enoto11,Archibold17},
these  regions responsible for the energy release and photon emission 
will change in their positions or beaming directions (Paper I).
This will generate some {\it phase noise} in their pulses,
which might have concentrated by chance at 55 ks and/or 27 ks.
However, the overall chance occurrence probability of  $\sim 2 \times 10^{-8}$,
derived in section~\ref{subsec:comparison_Suzaku},
remains still valid even considering
systematic errors caused by the phase noise,
because the phase-randomization process 
employed in section~\ref{subsec:significance}
should conserve the phase noise, assuming it to be approximately white.
Therefore, it is difficult to attribute the observed results
to an  accidental concentration  of the phase noise.
  
\subsection{Interpretation in terms of free precession}
\label{subsec:free_precession}
As argued above, the HXC pulse-phase modulation is
regarded as a regular phenomenon,
arising in an isolated NS which can be treated, 
on the relevant time scales,  as a rigid body.
This brings us back to our original idea 
of free precession (section~\ref{sec:intro}),
because it is the simplest and the most basic regular behavior
(except the rotation) of an axi-symmetric rigid body.
Specifically, the pulses are expected to show phase modulation
at the slip period $T$ of equation~(\ref{eq:slip}),
when the following three symmetry-breaking conditions are all satisfied (Paper II);
the star should be deformed to have $\epsilon \ne 0$ in equation (\ref{eq:epsilon});
the wobbling angle  (section~\ref{sec:intro}) should be $\alpha \ne 0$
as will be spontaneously realized when the deformation is prolate ($\epsilon>0$);
and the HXC emission pattern should be
asymmetric around the star's symmetry axis.
We hence infer that 
the NS in 4U~0142+61 keeps  free precession
with $\epsilon \ne 0$ and $\alpha \ne 0$ kept constant,
as neither would vary in several years,
but  the asymmetry of the HXC emission pattern changed considerably,
so that it was larger in 2009 and 2013
while smaller in the NuSTAR observation and in 2007.

\subsubsection{Assumed configuration}
\label{subsubsec:configuration}
\begin{figure*}[htb]
  \begin{center}
\includegraphics[width=16cm,]{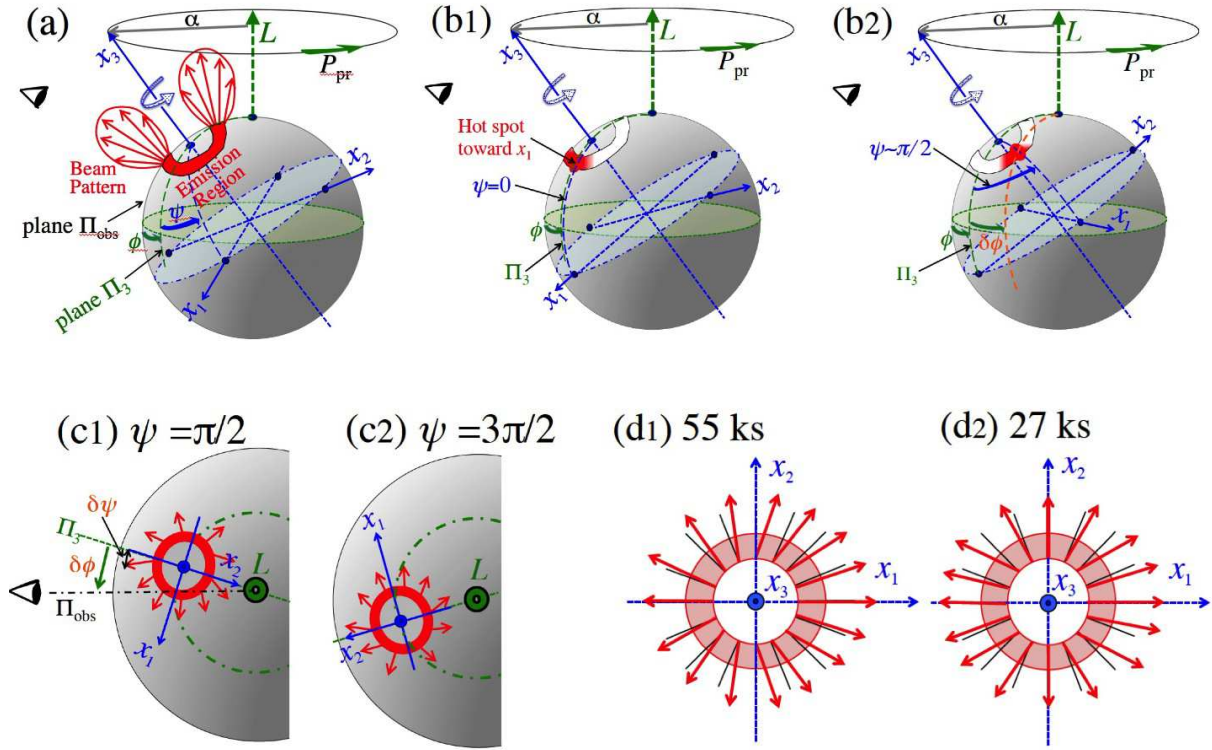}
 \end{center}
  \caption{
 A toy model to explain the pulse-phase modulation phenomenon of 4U~0142+61.
 (a) The assumed basic geometry of an axi-symmetric star,
 where $\vec{L}$ is the angular-momentum vector fixed to the inertial frame,
$(\hat{x}_1,\hat{x}_2,\hat{x}_3)$ denote the principal axes of the body,
$\Pi_{\rm obs}$ (=the plane of the sheet)
the plane containing $\vec{L}$ and the observer's line-of-sight,
$\Pi_{\rm 3}$  that defined by $\vec{L}$ and $\hat{x}_3$,
 $\alpha$ the wobbling angle, 
 $\phi$ the pulse phase,
 and $\psi$ the slip phase; see text for details.
A ring-shaped emission region in red is assumed,
together with a sketch of the beaming pattern.
(b1) The case wherein the ring-shaped region has a hot spot at
an azimuth of $+\hat{x}_1$, and the slip phase is $\psi=0$.
(b2) The same as panel (b1) but for $\psi \sim \pi/2$.
(c1) A top view onto $\vec{L}$, 
illustrating the case when the beaming direction is tilted along the ring
just one cycle per $2\pi$, and the slip phase is $\psi \sim \pi/2$.
(c2) The same as panel (c1), but for $\psi \sim 3\pi/2$.
(d1) The same beaming patterns as in panels (c1) and (c2),
but drawn on a plane perpendicular to $\hat{x}_3$,
where arrows indicate the projected beaming direction.
The tilt changes one cycle over the $2\pi$ azimuth.
The black ticks indicate the radial directions.
(d2) The same as panel (d1), but the tilt angle changes two cycles.
  }
\label{fig:toy_model}
 \end{figure*}

Along the above consideration, 
together with Paper I and Paper II, 
let us  consider a configuration 
 illustrated in figure~\ref{fig:toy_model}a.
There, the plane of the sheet, to be named $\Pi_{\rm obs}$,
is chosen to contain the angular momentum vector $\vec{L}$
fixed to the inertial frame, 
and the  line of sight of the observer located to the left.
The three principal axes of the NS are 
denoted as $(\hat{x}_1, \hat{x}_2, \hat{x}_3)$,
with $\hat{x}_3$ being the axis of symmetry 
which we identify with the dipolar magnetic axis.
The plane defined  by $\vec{L}$ and $\hat{x}_3$ 
will be denoted as $\Pi_{\rm 3}$,
and the angle from $\Pi_{\rm obs}$ to $\Pi_{\rm 3}$ as $\phi$.
Since $\Pi_{\rm 3}$ rotates with respect to $\Pi_{\rm obs}$ 
at the period $P_{\rm pr}$,
the observed pulse phase is identified with $\phi$
(item 2 of section ~\ref{sec:intro}).

The angle of $\hat{x}_1$ measured from $\Pi_{\rm 3}$ 
shall be denoted as $\psi$.
It is identical to $2\pi t/T $ in equation~(\ref{eq:modulation}),
with $\psi_0$  just its initial value,
and changes with the period $T$ to  specify  {\it slip phase}.
Assuming a prolate deformation ($\epsilon>0$; Paper I),
the precession is prograde,
so $\psi$ increases in the same direction as $\phi$,
though some $10^4$ times more slowly [equation (\ref{eq:slip})].

In a double-folding map like figure~\ref{fig:visual}a,
we can identify   $\phi$ and $\psi$
with $2\pi k/K$ (abscissa) and $\psi= 2\pi j/J$ (ordinate), respectively,
except a constant initial value.
In addition, $(\phi, \alpha, \psi)$ provide the three Euler angles 
to describe  the NS's attitude with respect to the inertial frame.

\subsubsection{Eission geometry for the  Suzaku results}
\label{subsubsec:emissinon_geometry_Suzaku}

In figure~\ref{fig:toy_model}a,   
a toy model for the hard X-ray emission is also illustrated.
There,  the  HXC is thought to be emitted from a ring-like source
(shown in red), located at one of the two magnetic poles.
The emission is assumed to be moderately collimated
towards the direction normal to the local NS surface.
We do not consider the emission from the other pole,
because its contribution, 
probably in the form of the secondary pulse peak,
is weak and free from the phase modulation effect (section~\ref{subsec:TwoPeaks}).
For the same reason, we do not require, either, 
that the 3rd symmetry breaking should take place 
simultaneously in both poles.

The source brightness is first assumed, as in panel (a),
to be constant along the ring azimuth,
and hence the emission  is symmetric around $\hat{x}_3$.
Then, the observed data are evidently insensitive to $\psi$,
and the pulsation will be exactly periodic  at  $P_{\rm pr}$.
Because the source is extended 
and the beaming is assumed to be not too strong, 
the pulse profile is expected to have a single broad peak at $\phi=0$,
if the line-of-sight makes a moderate inclination to $\vec{L}$.
(If the inclination is more pole on,  the pulse peak could split into two.)
A double-folding map such as figure~\ref{fig:visual}a
will be characterized by a straight  vertical ridge.
This case is considered to explain the SXC  pulses,
which are free from phase modulation.

Next,  assume that the source brightness is non-uniform along the ring,
and is concentrated, as in panels (b1) and (b2),
to form a hot spot towards the $+\hat{x}_1$ direction.
The pulse profile will become sharper than in panel (a),
because the source size has effectively diminished.
At the slip phase $\psi=0$ (panel b1),
the pulse peak will still reach us at $\phi=0$ with no phase shifts,
because the hot spot then lies on $\Pi_3$.
In contrast, at the slip phases $\psi \sim \pi/2$
when the configuration is as in  panel (b2),
the hot spot will cross $\Pi_{\rm obs}$ slightly
before $\Pi_3$ coincides with $\Pi_{\rm obs}$.
The pulse peak will hence arrive 
by a small amount $\delta \phi$  earlier than the expected  $\phi=0$.
The pulse-phase shift will return to 0 at $\psi \sim \pi$,  
and at $\psi \sim 3\pi/2$,
the peak will be delayed by $-\delta \phi$.
Thus, the pulse phase becomes dependent on $\psi$.

Supposing that $\alpha$ takes an intermediate value (e.g., $\sim 45^\circ$),
and the assumed ring has a relatively small half-cone angle $\beta$
(say, $\beta \sim 10^\circ$),
the amplitude of pulse-phase modulation would be at most 
$A/T = \delta \phi /2 \pi  \lesssim \beta/360^\circ \sim 0.03$.
However, $A$ can become larger
if the direction of the maximum emissivity,
or beaming direction,
is titled outwards from the local normal to the NS surface,
so that it points away from $\hat{x}_3$.
This is the configuration considered in Paper I, 
to explain the sinusoidal pulse-phase modulation 
with a relatively large amplitude of $A/T \sim 0.08$
as measured with the Suzaku HXD in 2009.

Strictly speaking, the pulse intensity will also be modulated,
to become maximum and minimum at $\psi=0$ and $\psi=\pi$, respectively,
although this effect was not clear with the Suzaku data 
due to the limited statistics (section~\ref{subsec:possible_origins}).
A search for this effect is one of our future tasks
(section~\ref{subsec:future}).

\subsubsection{Eission geometry for the  NuSTAR results}
\label{subsubsec:emissinon_geometry_NuSTAR}

So far, non-uniform {\it brightness} distributions
of the ring-shaped emission source,
possibly augmented by a tilt in the beaming direction,
were invoked to break the symmetry  around $\hat{x}_3$, or equivalently, in  $\psi$.
Although this has been successful on the 2009 and 2013 Suzaku data,
it may not afford adequate explanation to the NuSTAR results.
In fact, trying to explain the emergence of Zp27,
the ring might be assumed to have  two symmetric hot spots,
at  azimuths towards $\hat{x}_1$ and $-\hat{x}_1$.
Then, on a double-folding map, 
the two hotspots would wiggle each one cycle per $T$,
crossing a common vertical line {\em in an anti-symmetric way}.
This would  modulate, twice per $T$,
the width of the vertical ridge in the plot,
but would not produce  such periodic displacements
of the brightness centroid,
twice per cycle, as seen in figure~\ref{fig:55_27.5}a.

As an alternative idea to break the symmetry around $\hat{x}_3$, 
we may assume that  the brightness is rather uniform,
but the beaming direction is tilted  from the local normal,
this time azimuthally in the $\pm \psi$ direction,
and this tilt angle depends on $\psi$ (Paper I).
As in panels (c1) and (c2), 
let us assume that the beaming direction is
most forward-tilted by $\delta \psi>0$  at the azimuth  $-\hat{x}_2$,
and most backward-tilted by  $-\delta \psi$ at $+\hat{x}_2$,
so that the overall emission axis
(usually parallel to $\hat{x}_3$) is inclined towards $\hat{x}_1$.
Then, at a slip phase of  $\psi \sim \pi/2$ (panel c1)
and $\psi \sim 3\pi/2$ (paned c2), 
the pulse peak will reach the observer slightly 
before and after the prediction ($\phi=0$), respectively.
The amount of this pulse-phase shift is 
expressed as $\delta \phi = f \delta \psi$,
were $f$ is a form factor ($0<f<1$)
which depends on the beaming strength
and  approaches unity if the beaming is very strong.
Since we expect 
$A/P  \sim \delta \phi/2\pi = f \delta \psi /2 \pi$,
a tilt by $\delta \psi \sim 30^\circ$ and a moderate value of $f \sim 0.2$
would be adequate to explain the observed amplitude of  $A/P\sim 0.02$.
This condition is though to explain Zp55 of the present NuSTAR data.

The above case assumed that the tilt angle of the beaming direction
changes by one cycle along the ring azimuth.
This condition is  reproduced in panel (d1),
in the form of projection onto the plane perpendicular to $\hat{x}_3$.
In contrast, panel (d2) depicts the case
wherein the beam tilt angle changes twice over the $2\pi$ azimuth,
so that the emission is stronger in the $\hat{x}_1$-$\hat{x}_3$  plane
than in the $\hat{x}_2$-$\hat{x}_3$ plane.
Then, the pulse-phase shift is expected
to have a period of $T/2$ instead of $T$,
in a qualitative agreement with the appearance of Zp27.
The relative azimuths of panel (d1) and panel (d2)
has actually been selected to meet the observed relation
between $\psi_0 \sim 60^\circ$ for Zp55 
and $\psi_0 \sim 300^\circ$ for Zp27 (table~\ref{tbl:Zp55Zp27}).
Finally, a configuration like panel (d2) may also 
bring about fine structures in the pulse profile,
and could explain the $N$-dependence of Zp27 
(item 5 in section \ref{subsec:summary}).

In summary, the assumed ring-shaped emission region model can explain,
by invoking non-uniform brightness distributions
and an outward tilt in the beaming direction,
the  relatively large values of $A/P$ 
found with  Suzaku in 2009 and 2013.
The model can also explain, in terms of  
radiation-beam tilt in azimuthal directions,
the NuSTAR (and possibly the 2007 Suzaku) results
with the very small $A/P$ ratio and the prominent Zp27 feature.
These two modes of symmetry-breaking effects are likely 
to be taking place more or less simultaneously,
because Zp27 was observed in the 2013 Suzaku data as well
(figure~\ref{fig:Suzaku0913}b).

\subsection{Comparison with 1E~1547$-$5408}
\label{subsec:1E1547}
To generalize the results from 4U~0142+61 
to the knowledge of the magnetars as a class,
comparison with 1E~1547$-$5408 is valuable,
because a very similar hard X-ray pulse-phase modulation
was detected at $T=36.0^{+4.5}_{-2.6}$ ks
with Suzaku from this fastest rotating magnetar (Paper II).
Here, let us briefly update our discussion made in Paper II.

The two magnetars both exhibit pulse-phase modulations,
with two  important similarities.
One is that the  asphericity implied by equation (\ref{eq:epsilon}) is similar, 
$\epsilon=1.6\times 10^{-4}$ for 4U~0142+61
and $\epsilon=0.6\times 10^{-4}$ for 1E~1547$-$5408,
both assuming $\cos \alpha \approx 1$.
The other is that the effect is seen only in the HXC,
and absent (or much weaker) in the SXC.
In contrast to these similarities, 
the two objects differ considerably in their values of $A/P$,
as 1E~1547$-$5408 has $A/P \sim 0.25$
(reaching a quarter pulse phase),
which is much larger than those of  4U~0142+6, $A/P \lesssim 0.08$.
Thus, the relative amplitude $A/P$ is likely to exhibit 
a variety of behavior,  among different magnetars,
as well as among different observations of the same objects.

In Paper II,  the 36 ks phase modulation of 1E~1547$-$5408 
was explained invoking  a fan-beam like emission pattern.
This is because its pulse profile was double-peaked,
and the phase modulation appeared primarily as
periodic intensity interchanges between the two pulse peaks.
However, this concept is not much different from 
the ring-shaped emission region assumed in figure~\ref{fig:toy_model},
because such a ring-shaped region is expected to have a {\em conical emission pattern},
which will become a fan-beam configuration as 
the outward tilt of the beaming direct gets larger.
This suggests a possibility of constructing a more generalized model 
that can explain the behavior of the two sources,
in terms of differences in the model parameters
including $\alpha$, the viewing inclination,
the ring half-cone angle $\beta$,  
and the outward/azimuthal tilts of the beaming directions.

\subsection{Future works}
\label{subsec:future}
Before concluding the paper,
we may touch on  several future works to be performed.
The most obvious one is a more systematic re-analysis 
of the three Suzaku data sets (2007, 2009, and 2013) of 4U~0142+61,
plus another shorter one (gross 80 ks) acquired  in 2011 September
as a follow-up to a transient brightening of this magnetar.

Although the toy model of figure~\ref{fig:toy_model} appears to explain 
the Suzaku and NuSTAR observations of 4U~0142+61
in a consistent manner, it is still rather primitive.
We clearly need to make it more quantitative,
via numerical modeling of the geometry 
and beaming properties of the assumed emission region,
to examine whether the model 
can actually explain the pulse-phase modulation in the data,
as well as the observed pulse profiles 
and possible $\psi$-dependent intensity changes.
Furthermore, as mentioned in section~\ref{subsec:1E1547},
we need to generalize the model  
so that it can account for
not only the four observations of 4U~0142+61,
but also the behavior  of 1E~1547$-$5408.
Through such a modeling, we will be able to obtain
some estimates on $\alpha$ of the two objects,
which remain at present  only coarsely constrained 
to be neither $\approx 0$ nor $\approx 90^\circ$ (Paper II).
These are also left for our future studies.

An even more important future work is to find physical grounds 
to the toy model of figure~\ref{fig:toy_model},
and investigate mechanisms
which can explain the switch between Zp55 and Zp27
(item 4 in section~\ref{subsec:summary}).
Similarly, the transition between the large-amplitude modulation seen in 2009/2013
and the small-amplitude cases  as in 2007/2014 need to be accounted for.
Here, we only mention that the HXC could be 
produced via a quantum-electro-dyncamical process 
called {\it photon splitting} (e.g., \cite{Enoto10}),
where some input gamma-ray photons keep splitting towards softer photons,
as they propagate along the MF lines
that are stronger than the critical field value of 
$B_{\rm c} \equiv m_{\rm e}^2 c^2 /e \hbar 
= 4.4 \times 10^{9}$ T $= 4.4 \times 10^{13}$ G;
here,  $m_{\rm e}$, $c$, $e$, and $\hbar$ are
the electron mass, the light velocity, the elementary charge,
and the Dirac constant, respectively.
Then, the phenomena considered in figure~\ref{fig:toy_model}
could emerge as a result of  incessant changes
in the magnetic field configuration near the magnetic poles,
where multi-pole structure may be involved.
Conversely, further studies of the present phenomenon
will give valuable clues to  the emission mechanism of the HCX.

Finally, we need to increase the number of magnetars
that exhibit the same phenomenon,
in addition to 4U~0142+61 and 1E~1547$-$5408.
This will allow us to examine wheter magnetars are all axially deformed,
and if so, how their $\epsilon$  (and hence the toroidal MF strengths)
depends on their parameters including
the age, the pulse period, and the dipole MF strengths.
At the same time, a comparison of the same object
at different intensity levels would be of great value,
because 4U~0142+61 remained at a relatively similar intensity level
through the four observations.

\section{Conclusion}
\begin{enumerate}
\item As confirmed consistently with the 2009 and 2013
Suzaku data sets, and those from NuSTAR obtained in 2014,
the 8.69-s hard X-ray pulsation of 4U~0142+61
is phase modulated at a period of 55 ks,
whereas its soft X-ray pulsation is not.

\item The modulation amplitude varis by a factor 
of 4--7 on time scales of months to years,
and sometimes the 2nd harmonic of the 55 ks periodicity appears
at about half that period.

\item  
The scenario proposed in Paper I  has been reconfirmed:
the pulse-phase modulation is interpreted as
caused by free precession of the neutron star,
which is deformed to $\epsilon \sim 1.6 \times 10^{-4}$,
possibly due to internal magnetic fields reaching $\sim 10^{16}$ G.

\item Major roperties of the pulse-phase modulation are understood
if the hard X-ray source has a certain geometry,
and exhibits a variable degree of asymmetry 
in its position and/or beaming direction
around the star's axis.

\item The similarity between 4U~0142+61 and 1E~1547.0$-$5408,
discussed in Paper II, remains valid.
Namely, both objects are likely to be axially deformed to $\epsilon \sim 10^{-4}$,
and undergo free precession.

\end{enumerate}

\section*{Acknowledgements}
The authors thank Toshio Nakano and Yoshihiro Furuta 
for their help in data analysis and participation in discussion.
This work was supported partially by the MEXT
Grant-in-Aid for Scientific Research (C),
Grant No. 18K03694.


\bigskip
\begin{thebibliography}{99}
\bibitem[Archibold et al.(2017)]{Archibold17}
Archibald, F., Kaspi, V.,  Scholz, P., Beardmore, A., Gehrels, N., \& Kennea, J. 2017,
\apj, 834, 163
\bibitem[Brazier(1994)]{Zn2_94}
Brazier, K. T. 1994, \mnras, 268, 709
\bibitem[Buccheri et al.(1983)]{Zn2_83} 
 Buccheri, R., et al. 1983, \aap, 128, 245
\bibitem[Coti Zelati et al.(2017)]{CotiZelati17}
Coti Zelati, F.,  et al. 2017, \mnras, 471, 1819
\bibitem[Cutler(2002)]{Cutler02} 
Cutler, C. 2002, Phys. Rev. D., 66,  084025 
\bibitem[Enoto et al.(2017)]{Enoto17}
Enoto, T., et al. 2017, \apjs,  231,  8
\bibitem[Enoto et al.(2011)]{Enoto11}
Enoto, T., Makishima, K., Nakazawa, K., Kokubun, M.,  Kawaharada, M., 
Kotoku, J., \&  Shibazaki, N. 2011, \pasj,  63, 387
\bibitem[Enoto et al.(2010)]{Enoto10}
Enoto, T., Nakazawa, K.,  Makishima, K., Rea, N.,  Hurley, K., \& Shibata, S. 2010, 
\apjl, 722, L162
\bibitem[Gavriil \& Kaspi (2002)]{4U0142_Bd}
Gavriil, F., \&   Kaspi, V.  2002, \apj 567, 1067
\bibitem[Grimani(2016)]{disk_precession}
Grimani, C. 2016, \mnras, 460, 2186
\bibitem[Ioka \& Sasaki(2004)]{Ioka+Sasaki04}
Ioka, K., \& Sasaki, M. 2004,  \apj, 600, 296
\bibitem[Kokubun \etal (2007)]{HXD2}
Kokubun. M., et al.\ 2007, \pasj, 59, S53
\bibitem[Kuiper et al.(2012)]{Kuiper12}
Kuiper, L.,  Hermsen, W.,  den Hartog, P. R., \& Urama, J. O. 2012, \apj, 748, 133
\bibitem[Makishima(2016)]{Max16} 
Makishima, K.  2016,
Proc. Japan Academy, Ser. B,  92, 135
\bibitem[Makishima et al.(2014)]{Makishima14} 
Makishima, K., Enoto, T., Hiraga, J. S., Nakano, T., Nakazawa, K.,
Sakurai, S., Sasano, M., \& Murakami, H. 2014, 
Phys. Rev. Lett., 112, 171102 (Paper I)
\bibitem[Makishima et al.(2016)]{Makishima16} 
Makishima, K., Enoto, T.,  Murakami, H., Furuta, Y., Nakano, T,
Sasano, M., \& Nakazawa, K. 2016, 
PASJ,  68S, 12 (Paper II)
\bibitem[Mereghetti(2008)]{Mereghetti08}
Mereghetti, S. 2008, \aapr, 15,  225
\bibitem[Takahashi \etal (2007)]{HXD1}Takahashi, T., et al.\ 2007, \pasj, 59, S35
\bibitem[Tendulkar \etal (2015)]{TEA15}
Tendulkar, S. P., et al. 2015, \apj, 808, 32
\bibitem[Thompson \& Duncan(1995)]{Magnetar} 
Thompson, C.,  \&   Duncan, R. C. 1995, \mnras,  275, 255
\bibitem[Zezas \etal(2015)]{fallback_disk} 
Zezas, A.,  Tr\"{u}mper, J. E., and Kylafis, N. D. 2015,
\mnras, 454, 3366-3375
\end{thebibliography}
\end{document}